\def\sx{\sigma^x}

\def\sz{\sigma^z}
\def\D{\Delta}
\def\e{\varepsilon}
\def\l{\lambda}
\def\mz{\mu_z}
\def\x1{\chi_z}
\def\x2{\chi_{zz}}
\def\tr{{\rm tr}}
\def\Q{\theta}
\def\w{\omega}
\def\w0{\omega_0}
\def\W{\Omega}
\def\W0{\Omega_0}
\def\nr{|n\rangle}
\def\nl{\langle n|}

\def\pr{|p\rangle}
\def\Pl{\langle p|}
\def\qr{|q\rangle}
\def\ql{\langle q|}

\documentclass[pra,aps,floatfix]{revtex4}

\usepackage{graphicx}

\begin{document}
\title{Nonlinear response and observable signatures of equilibrium entanglement}

\author{A.M.~Zagoskin\footnote{zagoskin@physics.ubc.ca}}
\affiliation{Department of Physics and Astronomy, The University of British Columbia, Vancouver, BC, V6T 1Z1, Canada}
\affiliation{Digital Materials Lab., Frontier Research System, RIKEN, 2-1 Hirosawa, Wako-shi, Saitama 351-0198, Japan}
\affiliation{Department of Physics, Loughborough University, Leicestershire, LE11 3TU, UK (permanent address) }

\author{A.Yu.~Smirnov}
\affiliation{Digital Materials Lab., Frontier Research System, RIKEN, 2-1 Hirosawa, Wako-shi, Saitama 351-0198, Japan}
\affiliation{CREST, Japan Science and Technology Agency, Kawaguchi,
Saitama, 332-0012, Japan} 

\author{S.K.~Gupta}
\affiliation{Department of Physics and Astronomy, The University of British Columbia, Vancouver, BC, V6T 1Z1, Canada}

\author{I.S.~Slobodov}
\affiliation{Department of Physics and Astronomy, The University of British Columbia, Vancouver, BC, V6T 1Z1, Canada}

\begin{abstract}
We investigate how equilibrium  entanglement is manifested in the nonlinear response of an $N$-qubit system. We show that in the thermodynamic limit the irreducible part of the $n$th-order nonlinear susceptibility indicates that the eigenstates of the system contain entangled $(n+1)$-qubit clusters. This opens the way to a directly observable multiqubit entanglement signature. We show that the irreducible part of the static cubic susceptibility of a system of four flux qubits, as a function of external parameters, behaves as a  global 4-qubit entanglement measure introduced in Ref.\cite{Love}. 

We discuss the possibility of extracting purely-entanglement-generated contribution from the general multipoint correlators in a multiqubit system. 
\end{abstract}

\maketitle

\section{Introduction}

The ``equilibrium", or ``thermodynamic", entanglement in
spin systems is a  quantification of the degree of non-classicality of the eigenstates   of the system  (e.g. \cite{Osborne2002,Osterloh2002,Martin2002,Ghosh2003,Vidal2003}). In quantum phase transitions it can be regarded as a specific order parameter \cite{Verstraete2004,Brandao2005}. It distinguishes between the classical correlations, produced by interaction, and the purely quantum ones. Both types of correlations are contained in  multi-point Greens' functions of the system. However,  there is no simple recipe to separate them. This problem is exacerbated by the absence of a convenient measure of entanglement  for a mixed state of more than  two subsystems.  Quantum state tomography can reconstruct all qubit states of a general spin system, but the number of required operations grows exponentially with $N$ \cite{Liu2004,Liu2005}.


Various entanglement measures were expressed through spin correlators in \cite{Glaser2003}. Observable ``entanglement witnesses" \cite{Wu2005} or ``entanglement estimators" \cite{Roscilde2004} were proposed as well. In general, a number of measurements is required to restore the measure of  entanglement in equilibrium, though not as large as in the case of quantum state tomography.

Equilibrium entanglement is particularly important in the context of adiabatic quantum computing (AQC) \cite{AQC}. From the equivalence of AQC and standard quantum computing \cite{AQC-standard,AQC-standard-2} and the polynomial equivalence of quantum and classical computing in case when there is no global entanglement\cite{Vidal} we   infer  that global entanglement of the ground state is necessary for an efficient AQC.  The equilibrium state of a  system is a sum of projectors on its eigenstates. It is therefore desirable to be able to check that at least some of these project on {\it globally entangled} eigenstates.

The investigation of equilibrium entanglement allows us to shift the focus from the state to the Hamiltonian (e.g. its entanglement-generation ability\cite{Bandyo2004}). This opens a new venue of investigation. Suppose the Hamiltonian of a quantum system can be restored without free parameters from a series of measurements (which means that the behaviour of the system, including level anticrossings etc, is quantitatively described under the assumption that it is in a thermal equilibrium mixed state produced by this Hamiltonian). Then all the eigenstates of the system can be determined, and their entanglement measures can be computed.

The experimental investigation of equilibrium entanglement along these lines was performed for the sytems of two\cite{Izmalkov2004} and four\cite{Grajcar2005,Love} flux qubits. It relied on continuous measurement of the magnetic susceptibility of an $N$-qubit system. The qubits were controlled locally; the global response was measured, and all the operations were slow on the scale of characteristic qubit times. 
The quantification of entanglement in both cases was done ``post-mortem" and showed a high degree of entanglement for the ground and first excited states.

Remarkably, the results of  \cite{Izmalkov2004} also contain a directly observable, ``in vivo" 2-qubit entanglement signature, so-called IMT deficit (a difference between the sum of signals from two qubits driven through their degeneracy points separately, and the signal when they pass the co-degeneracy point simultaneously; the abbreviation comes from the measurement technique employed). It appears in the linear susceptibility, which is proportional to 
\begin{equation}
R_{pq} \sim \sum_{i,j}\Pl{\sz_i}\qr\ql{\sz_j}\pr,\:\: p\neq q. \label{eq_Rpq}
\end{equation}
Here $\pr,\qr$ are the eigenstates of the Hamiltonian, and $\sz_j$ is proportional to the magnetic moment of the $j$th qubit in $z$-direction. (The actual formula (e.g. Eqs.(2,3) of Ref.\cite{Izmalkov2004}; see also \cite{Smirnov03}) takes into account asymmetries in couplings of qubits to the external field source; while necessary for the quantitative match to the experiment, it is not conceptually important.) If the eigenstates are not entangled (that is, do not contain components differing by two qubit flips in the $z$-basis), part of this term (with $i\neq j$) disappears. The only requirement for the observation of this signature is, as we stated above, the ability to manipulate qubits separately and measure the total response of the system.

The linear response measurements will not provide a   directly observable  signature of equilibrium entanglement with $N>2$. The expressions like (\ref{eq_Rpq}) arise  in the lowest-order perturbation theory and therefore contain only first-order commutators between the qubit operators (coming from the expression for the observable, and from the perturbation Hamiltonian). Therefore $R_{pq}$ can only catch the absence of 2-tangled components, but is insensitive to the higher-order (e.g. GHZ-type) states. This is why the appearance of globally entangled eigenstates  ($|\uparrow\downarrow\uparrow\downarrow\rangle \pm |\downarrow\uparrow\downarrow\uparrow\rangle$) in \cite{Grajcar2005} could be only inferred from the ``post mortem" analysis.

In order to obtain information about entanglement with $N>2$ without reverse-engineering of the Hamiltonian, one requires $N$-point  correlators. In classical case, they can be extracted  from the noise fluctuations (``noise of the noise"), but it is not clear yet how and whether this can be generalized to the quantum case. A more immediate approach is to measure {\em nonlinear} response of a multiqubit system. In practice, at least quadratic and qubic effects should be observable, thus providing signatures of 3- and 4-qubit  equilibrium entanglement.

\section{Entanglement and higher-order correlators}\label{sec2} 

First consider an $N$-point correlator of single-qubit spin operators, $S_k(t_k)$, acting on qubit $k$ at the moment $t_k$,
\begin{eqnarray}
 C_N(\{k, t_k\},\pr) = \Pl S_1(t_1)S_2(t_2)\dots S_N(t_N)\pr, \:k=1.\dots N,\label{eq_C}
\end{eqnarray}
where $\pr$ is some eigenstate of the Hamiltonian. 
The sequence of operators in (\ref{eq_C}) generally flips $N$ spins in the state $\pr$. Therefore the correlator disappears unless $\pr$  is a superposition of components differing by $N$ spins.
 In general case, one expects that $C_N$ should disappear or at least diminish in the absence of entangled states involving at least $N$ qubits.

 \subsection{Equilibrium $N$-qubit  entanglement} 
 
\begin{figure}[h]
\includegraphics
[width=8.8cm]{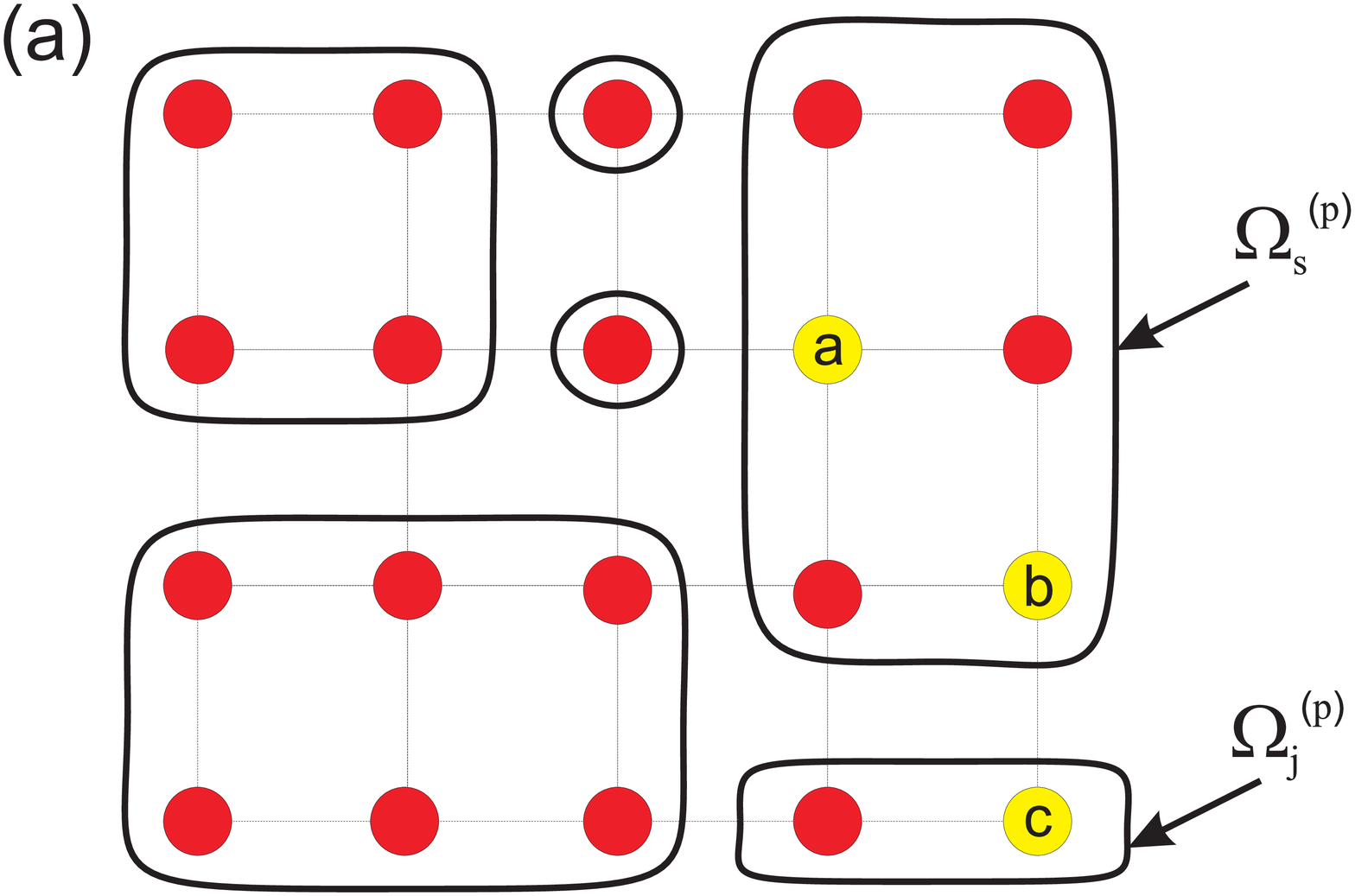}
\includegraphics
[width=8.8cm]{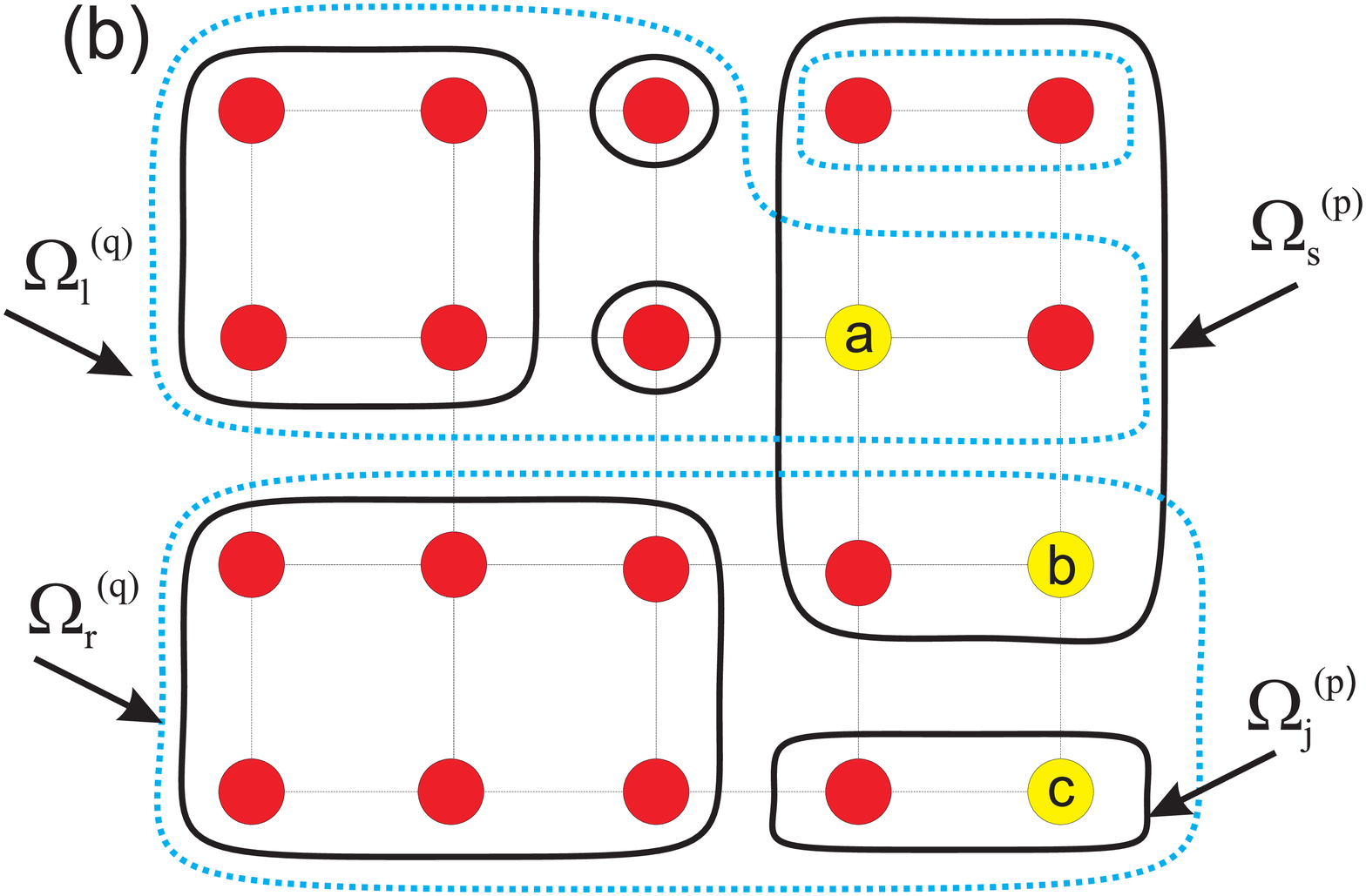}\\
\includegraphics
[width=8.8cm]{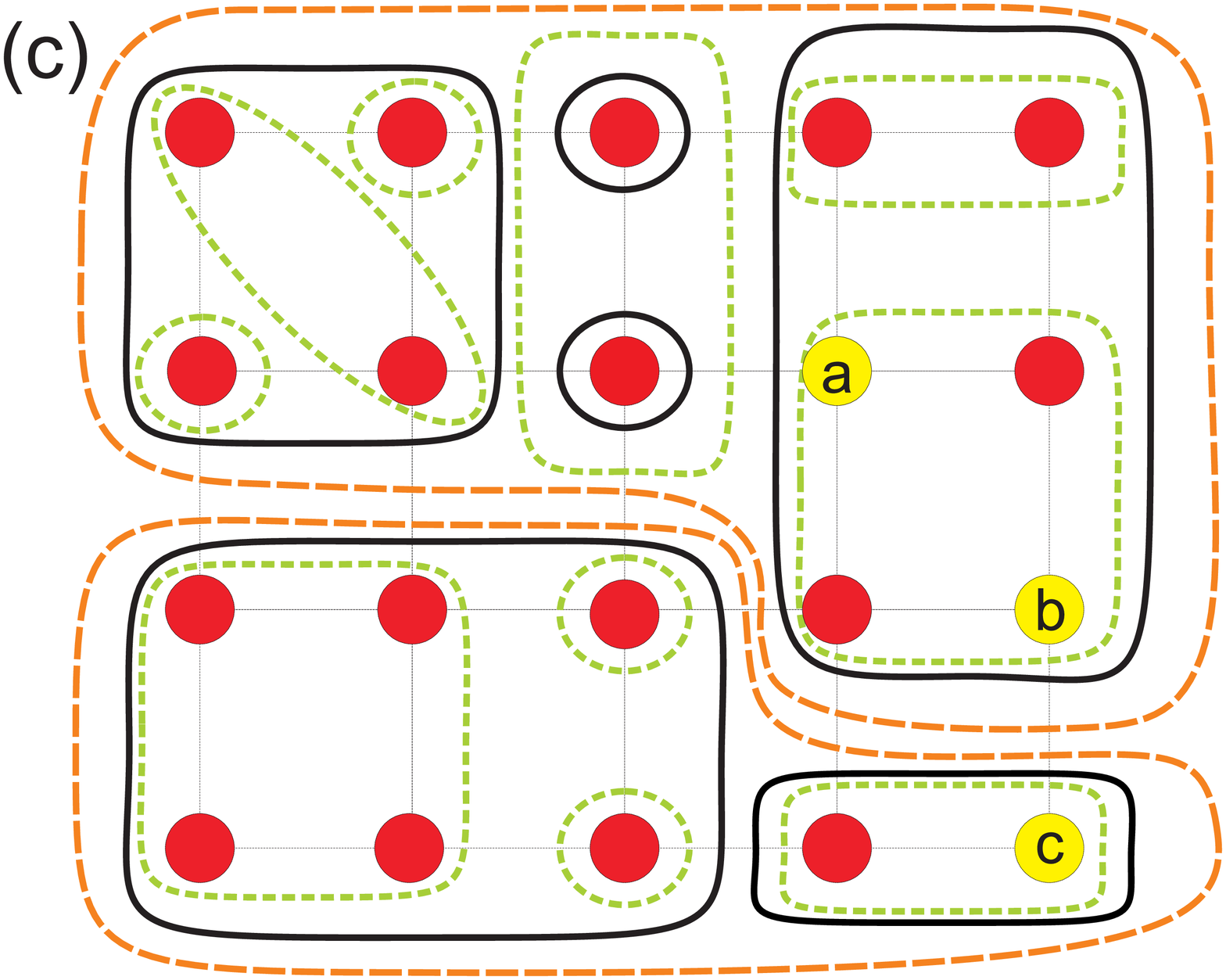}
\includegraphics
[width=8.8cm]{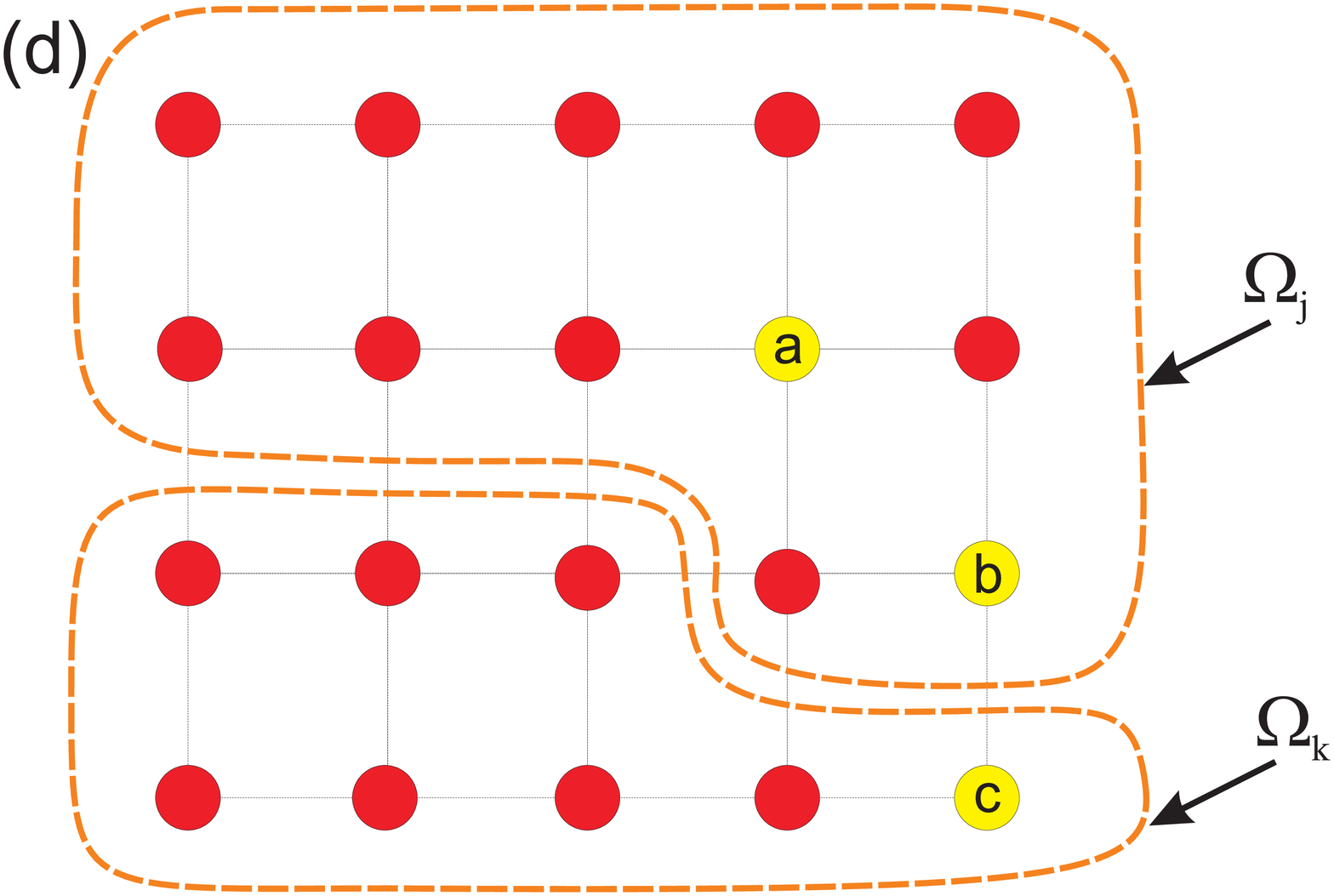}
\caption{(Colour online.) Russian-doll condition and partitioning the system into maximal clusters. (a) An eigenstate $\pr$ of an $M$-qubit system  can be presented as a product of functions, which depend only on states of qubits belonging to the clusters $\Omega_1^{(p)},\:\Omega_2^{(p)},\dots$. 
In other words, an eigenstate determines the partition of the system 
 into mutually non-entangled clusters. (b) Two sets of qubit clusters, for eigenstates $\pr$ and $\qr$, are shown. The Russian-doll condition is not satisfied, since qubits $a$ and $b$ are entangled in $\pr$, but not entangled in $\qr$, while qubits $b$ and $c$ are not entangled in $\pr$, but entangled in $\qr$. (c) Russian-doll condition is satisfied for three eigenstates. The entangled clusters for them are marked by solid, dotted and dashed lines respectively.  The   largest "Russian dolls"  are the maximal clusters (clusters containing the largest number of qubits, mutually entangled in some eigenstate of the system). (d) The unique partition of the system of Fig.(c) into maximal clusters (clusters $\Omega_j$ and $\Omega_k$).  The system is globally entangled if a  maximal cluster contains all qubits.  }\label{Fig1-clusters}
\end{figure}

In a system of $M$ qubits, every eigenstate of the Hamiltonian can be written as a product of states defined on qubit clusters $\Omega_j^{(p)}$ of no more than $M$ qubits each (Fig.\ref{Fig1-clusters}a), \begin{equation}
\pr = \prod_{\Omega_j^{(p)}} |p_{\Omega_j^{(p)}}\rangle,\:\:\: \bigcup_j \Omega_j^{(p)} = {\cal M},
\end{equation}
in such a way that the qubits from different clusters are not entangled. Here ${\cal M}$  is the set of all qubits in the system; $|{\cal M}|=M$. Since every time we deal with a pure state, this operation does not produce any problems. The complement of a cluster $\Omega$ we denote by $\bar{\Omega} \equiv {\cal M} \backslash \Omega$.
 
For the following, we need to impose on the Hamiltonian the \\
\\
\noindent{\bf Russian-doll condition} (see Fig.\ref{Fig1-clusters}): If in any eigenstate $|p\rangle$ qubits $a,b \in \Omega_s^{(p)}$, while qubit $c \notin \Omega_s^{(p)}$, then there is no such eigenstate $|q\rangle$  and cluster $\Omega_r^{(q)}$, that $a \notin \Omega_r^{(q)}$, while $b,c \in \Omega_r^{(q)}$.\\
Equivalently, it can be stated as follows: if any two clusters $\Omega_s^{(p)},\:\Omega_r^{(q)}$ have a common qubit, then one of these clusters must be contained in the other (like Russian dolls):
\begin{equation}
\Omega_s^{(p)}\,\cap\,\Omega_r^{(q)} \neq \oslash \,\Rightarrow\, \left(\Omega_s^{(p)}\,\subseteq\,\Omega_r^{(q)}\right)\vee\left(\Omega_r^{(q)}\,\subseteq\,\Omega_s^{(p)}\right).
\end{equation}

The Russian-doll condition allows us to {\em uniquely} partition the $M$-qubit system in equilibrium into {\em mutually non-entangled} clusters ({\em maximal clusters}) $\Omega_j$, \begin{equation}{\cal M} =  \bigcup_j \Omega_j; \:\: \Omega_j = \max_p(\Omega_j^{(p)}),\end{equation}
each containing no more than $M$ qubits (see Fig.\ref{Fig1-clusters}c,d). 
This is a physically plausible picture, and we conjecture that most $N$-qubit  Hamiltonians  satisfy the Russian-doll condition. Establishing the limits of applicability of this conjecture will be the subject of further research.

In a system satisfying the Russian-doll condition,   the existence of a $k$-qubit maximal cluster is equivalent to any state of the system being   $k$-producible (see Ref.\onlinecite{producibility}. 
We can therefore say that equilibrium $N$-qubit  entanglement  
exists in a system of $M \geq N$ qubits, if at least one maximal cluster contains at least $N$ qubits. 
Indeed,  then in equilibrium the density matrix of the system is a sum of projectors on eigenstates, at least one of which supports $N$-qubit entangled clusters.

 Our goal is to find an observable signature  of equilibrium $N$-qubit entanglement, that is, an observable physical quantity which will be zero unless the system contains maximal clusters of size $N$. Let us check the handwaving arguments we made around Eq.(\ref{eq_C}) and see whether certain $N$-point correlators can be used this way.

 \subsection{Analysis of irreducible correlators of one-qubit operators}
 
Returning to (\ref{eq_C}), one can easily express it as a sum over the internal labels $p_2,p_3,\dots p_N$ of the correlators ($p_1 \equiv p$)
\begin{eqnarray}
 c_N(\{k,   |p_k\rangle\}) = \langle p_{1}| S_1 |p_2\rangle\langle p_2|S_2|p_3\rangle\dots \langle p_N| S_N |p_1\rangle,\:k=1.\dots N, \label{eq_c}
\end{eqnarray}
with the appropriate energy denominators, similar to (\ref{eq_Rpq}). (This is done by the repeated insertion in (\ref{eq_C}) of the closure relation, $\hat{\bf 1} = \sum_p \pr\Pl$.)

Now let us concentrate on the {\em irreducible correlators}, that is, the ones where no two states $|p_i\rangle, |p_j\rangle, i\neq j$ coincide.  The term reflects the following property of the correlator (\ref{eq_c}). Obviously from the definition, $c_N(\{k,   |p_k\rangle\})$ is  invariant with respect to the cyclic transpositions of the labels $p_k$. If some of these labels appear more than once, then $c_N(\{k,   |p_k\rangle\})$  is also invariant with respect to shorter cyclic transpositions in substrings, which end in the identical $p_k$'s.

For an irreducible correlator $c_N^{\rm irr}(\{k,  p\})$ there are two possibilities: either all $N$ qubits, on which act the operators $S_j$, belong to the same maximal cluster $\Omega$, or   they belong to  different maximal clusters.    

\subsubsection{Same maximal cluster}

In this case   the states of the system can be written as $\nr = \nr_{\Omega}\otimes\nr_{\bar{\Omega}}$ etc, where $\bar{\Omega}$ is the complement of $\Omega$: $$\Omega\,\cup\,\bar{\Omega}\,=\,{\cal M}.$$ (This follows from the definition of entanglement and the fact that in any eigenstate of the Hamiltonian $\Omega$ and $\bar{\Omega}$ do not contain mutually entangled qubits.) We therefore obtain 
\begin{equation} 
c_N^{\rm irr}(\{k,   p\}) =   
(\cdots )\delta_{p_1p_2}(\bar{\Omega})\delta_{p_2p_3}(\bar{\Omega})\cdots\delta_{p_Np_1}(\bar{\Omega}); \label{eq_A}
\end{equation}
(the Kronecker symbol $\delta_{pq}(\Omega)$ means that the states $p$ and $q$ coincide on the set $\Omega$). Eq.(\ref{eq_A}) reflects the fact that the states of the qubits {\it outside}  $\Omega$ are not changed by the operators acting on qubits inside $\Omega$, and can therefore only differ on $\Omega$. The number of states which only differ on a cluster of size $|\Omega| \equiv N_{\Omega}$ is $2^{N_{\Omega}}.$ Therefore there can be no irreducible correlators with $N>2^{N_{\Omega}}$ (otherwise the indices must repeat). Of course, this is not a serious limitation.

\subsubsection{Different maximal clusters}

\begin{figure}[h]
\includegraphics[width=8.8cm]{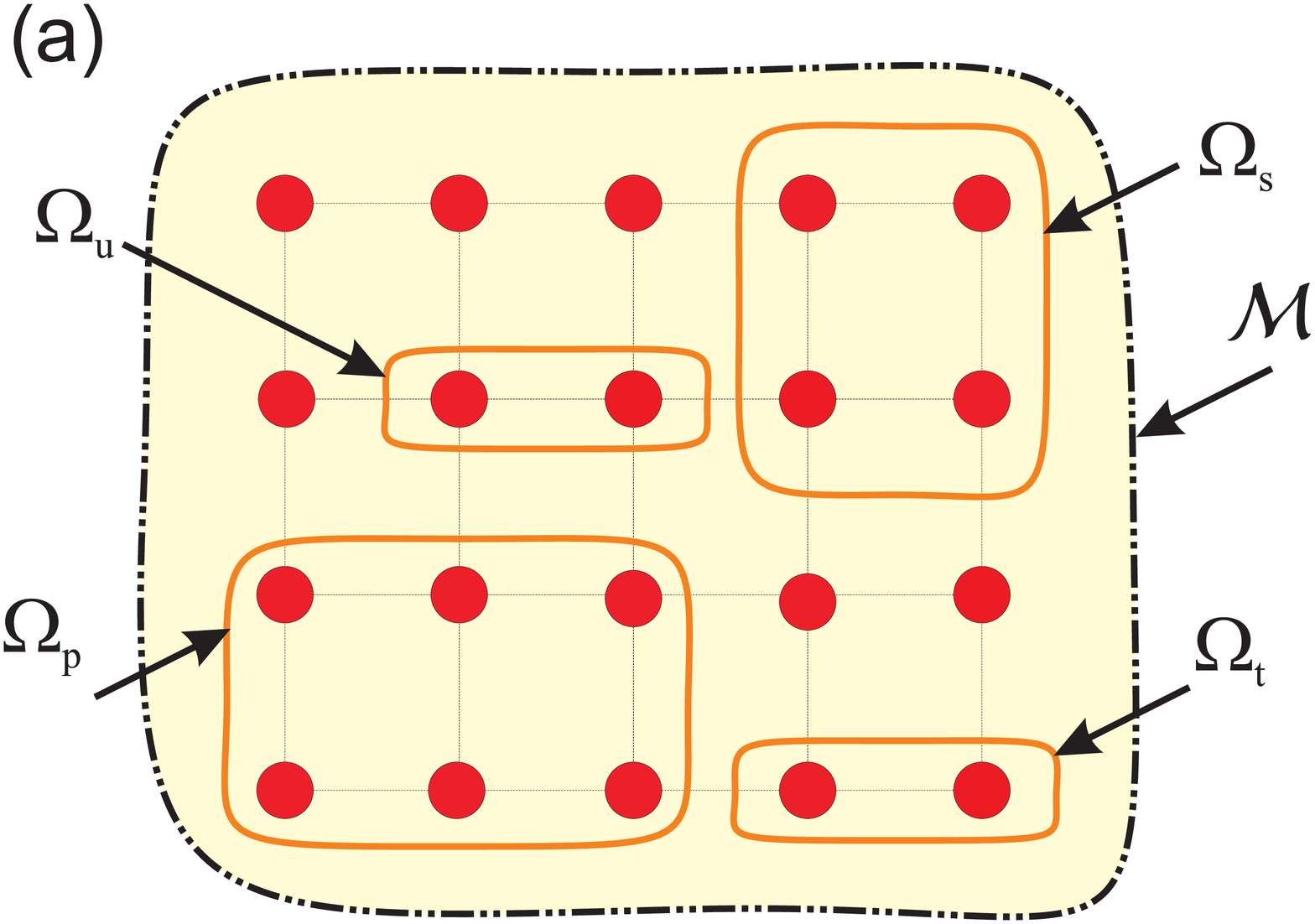}\includegraphics[width=8.8cm]{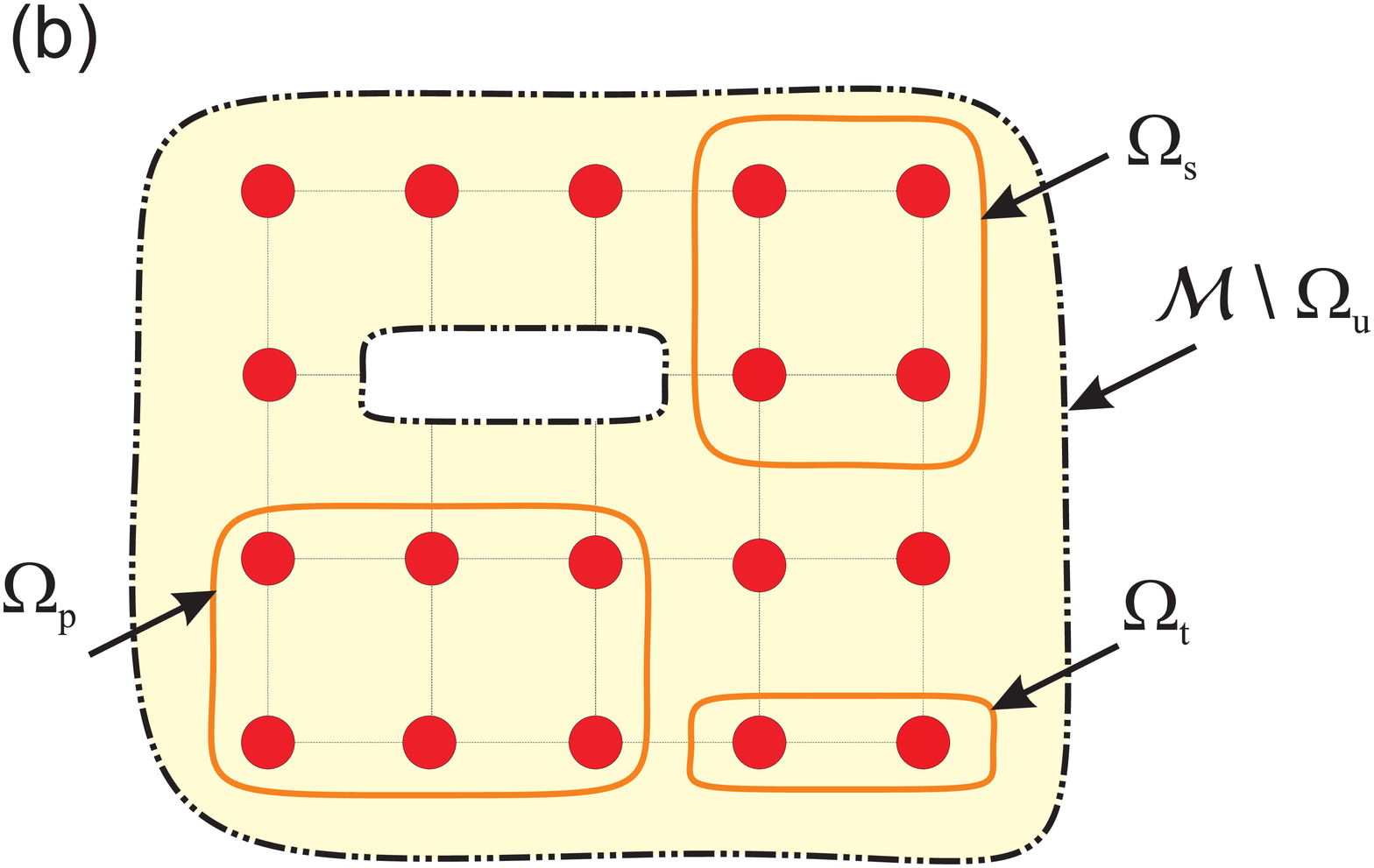}\\
\includegraphics[width=8.8cm]{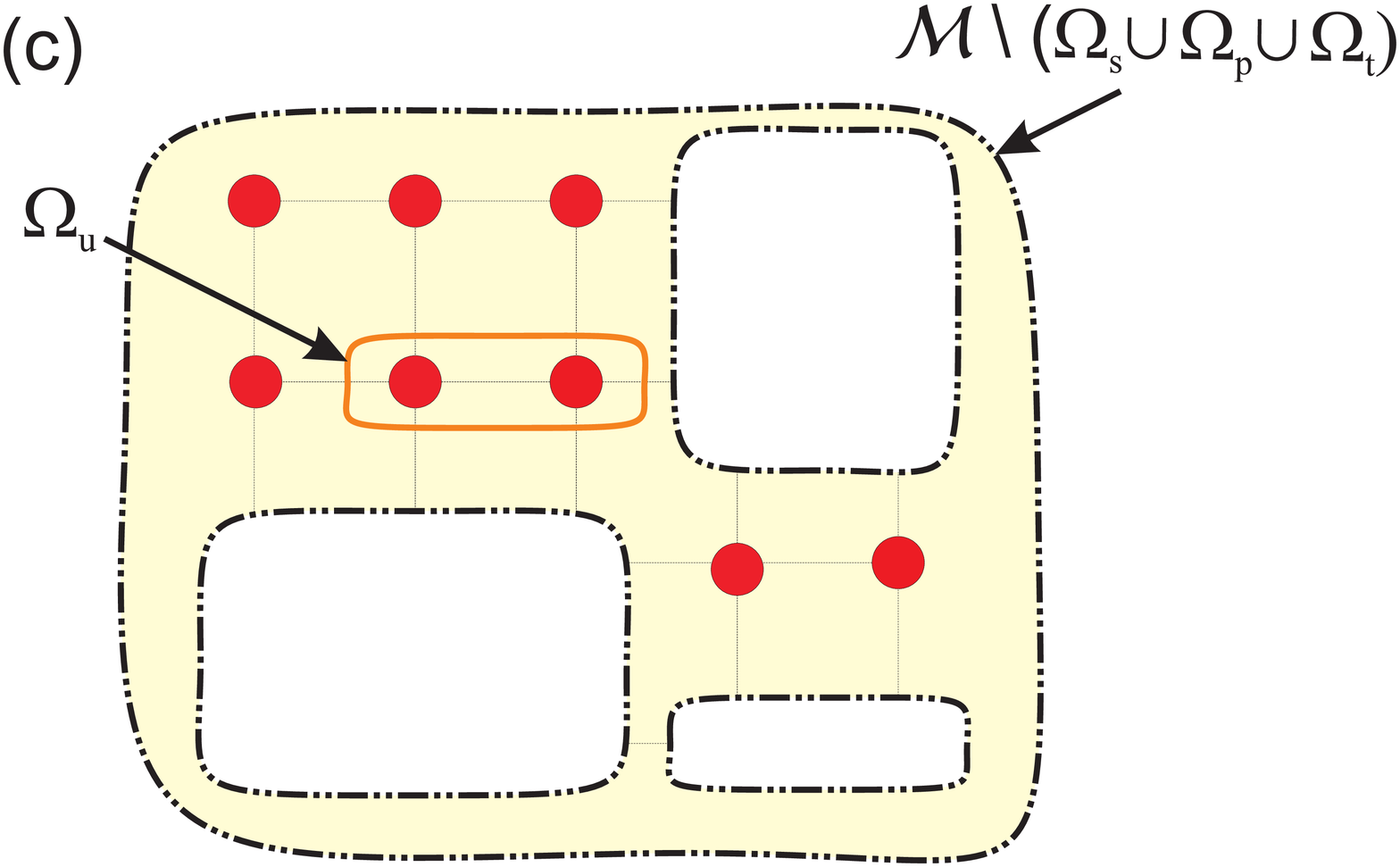}\includegraphics[width=8.8cm]{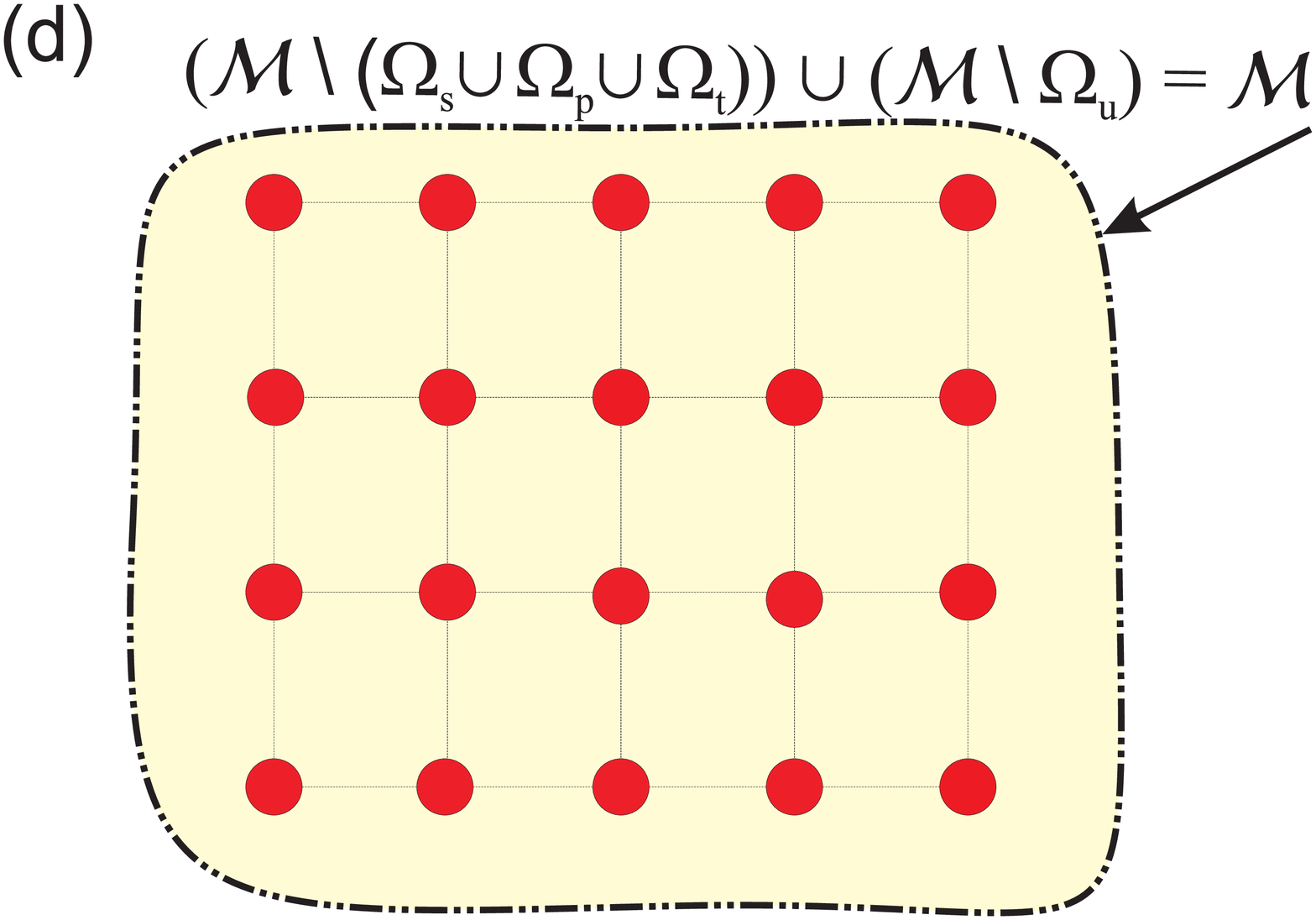}\\
\includegraphics[width=8.8cm]{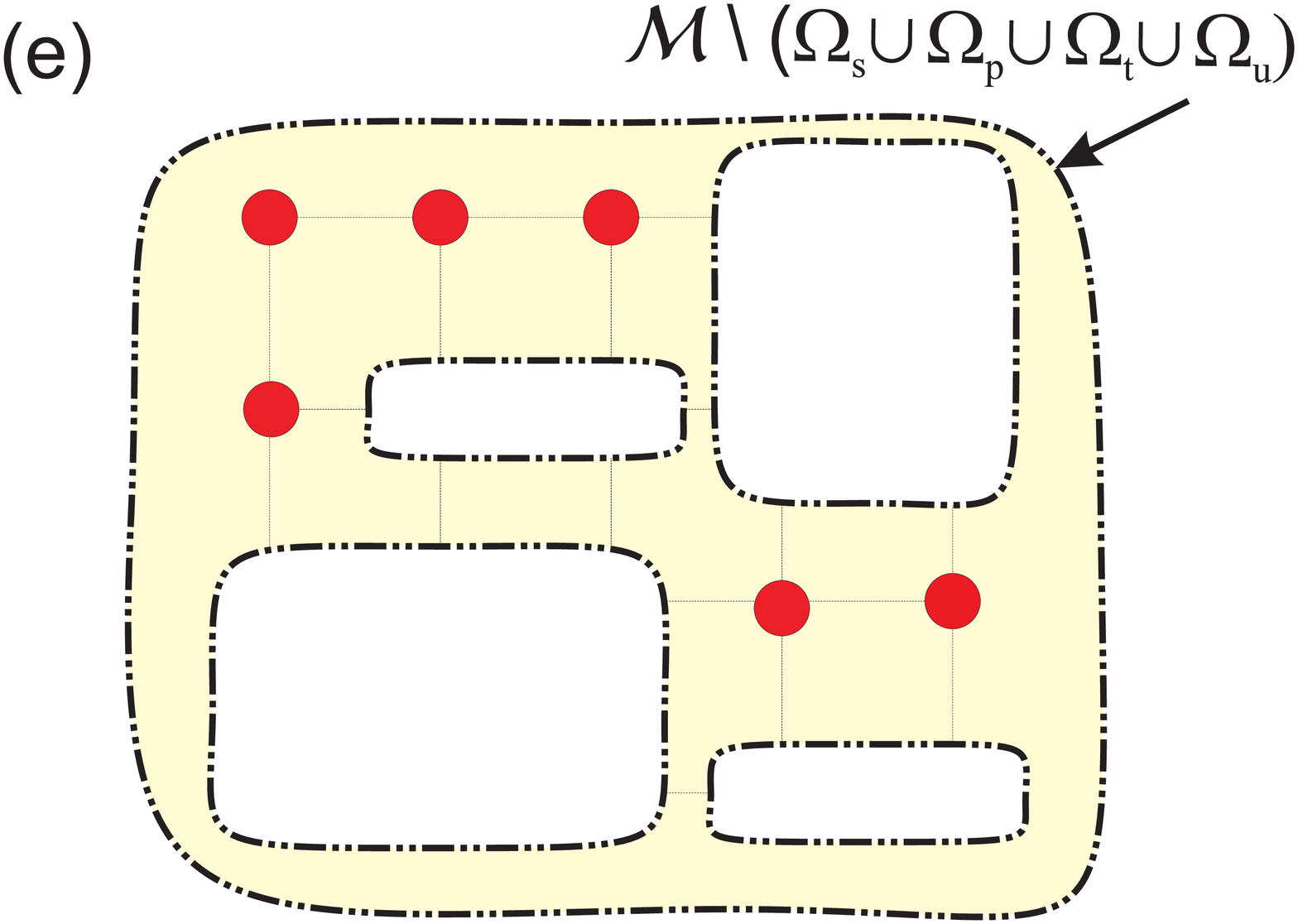}\includegraphics[width=8.8cm]{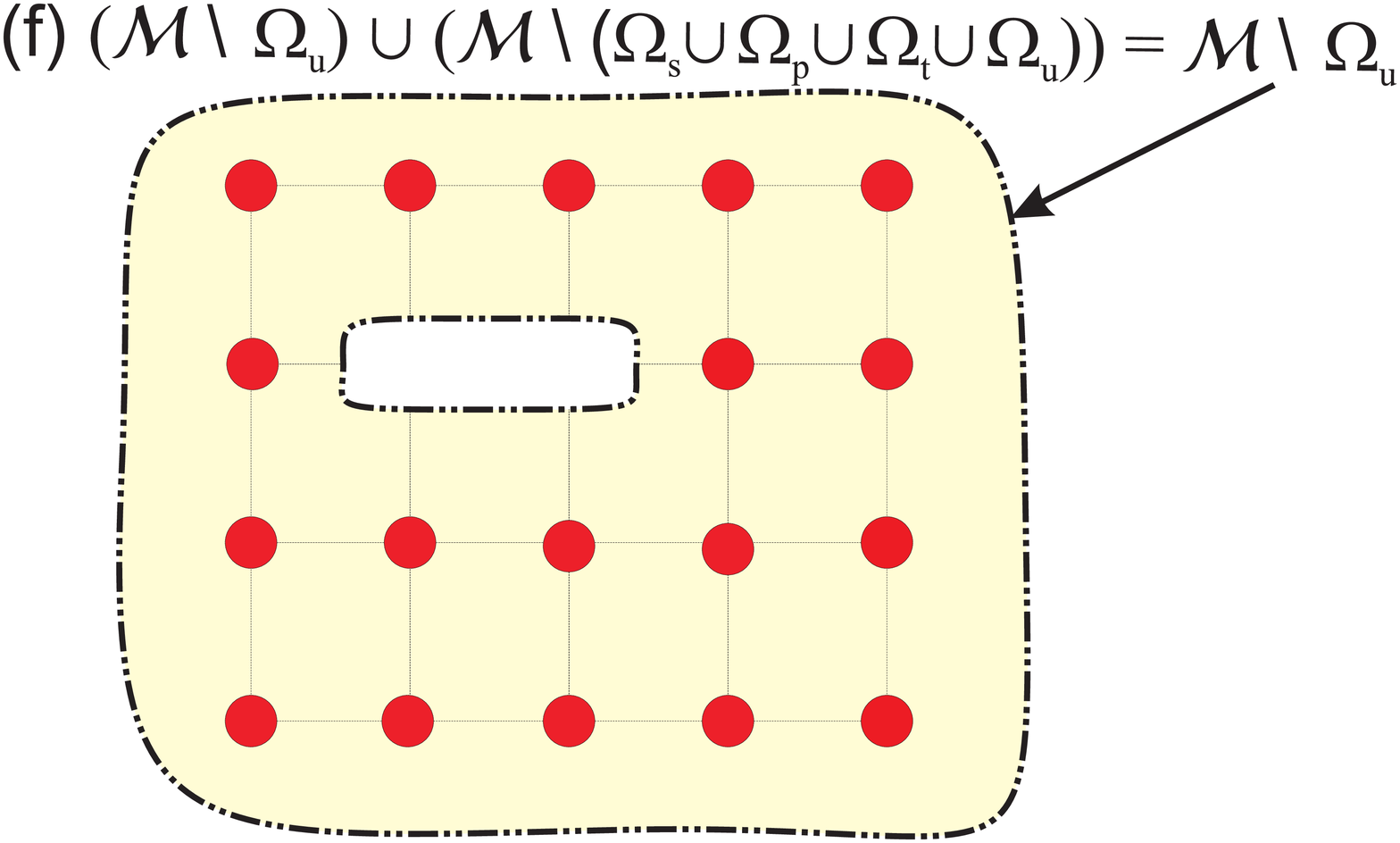}
\caption{(Colour online.) Joint and disjoint irreducible correlators. (a) The set of all qubits, $\cal M$, and maximal clusters called by the index string $S$. (b) The complement ${\cal M}\backslash\Omega_u \equiv \bar{\Omega}_u$, of the cluster $\Omega_u$.  (c) If $\Omega_u$ is unique, then the complement  of the sum of all other clusters called by the string $S$, ${\cal M} \backslash (\Omega_s \cup \Omega_p \cup \Omega_t)$, contains $\Omega_u$.  (d) Therefore the sum of (b) and (c) is the set of all qubits, $\cal M$. (e) If $\Omega_u$ is not unique, the  rest of the called clusters contains $\Omega_u$, and its complement does not. (f) Then the sum of the complements, (b) and (e), is not $\cal M$, but only $\bar{\Omega}_u$ (f). }\label{Fig2-strings}
\end{figure}

Now the string of indices $$S = 'j_1j_2\dots j_N',\:$$ which labels the qubit operators, consists of $1 < R \leq N$ substrings,  $S = s_1s_2\dots s_R.$  
In each substring the indices belong to the qubits in the same cluster, and there are $R$ such clusters involved: $\Omega_1, \Omega_2, \dots \Omega_R$ (see Fig.\ref{Fig2-strings}a).
Some of  these clusters may coincide (i.e. called more than once), with the only trivial restriction:
Clusters corresponding to adjacent index substrings, e.g. $\Omega_r$ and $\Omega_{r\pm 1}$, must be different; otherwise these  substrings concatenate.

The $r$th substring (the indices in which run from  $j^r_1$ to $j^r_{N_r}$, and $j^r_{N_r}+1 \equiv j^{r+1}_{1}$)  yields the product of the Kronecker symbols, ensuring that on $\bar{\Omega}_r$ (the complement of $\Omega_r$) 
\begin{equation}
|p_{j^r_1}\rangle = |p_{j^r_1+1}\rangle = \dots = |p_{j^r_{N_r}+1}\rangle \equiv |p_{j^{r+1}_{1}}\rangle. 
\end{equation}

To shorten the presentation, we introduce the following definitions:\\
\noindent
{\bf Definition 1:}  Substrings are {\em mutual}, if they contain qubit indices from the same maximal cluster.\\
\noindent
{\bf Definition 2:} A {\it unique} substring is the substring, which has no mutuals in $S$. \\
\noindent
{\bf Definition 3:} A {\em joint} correlator is the correlator, the index string of which contains at least one unique substring. The index string of a {\it disjoint} correlator does not contain unique substrings.

Now consider a joint correlator with  the unique substring $s_u$ (Fig.\ref{Fig2-strings}b,c): $$S = s_1 s_2 \dots s_{u-1} s_u s_{u+1} \dots s_{R-1} s_R.$$ 
Then, from Eq.(\ref{eq_c}) we obtain:
  \begin{eqnarray}
c_N^{\rm joint} = (\cdots)\: \delta_{p_1p_{j^2_1}}\left(\bar{\Omega}_1\right)\cdots
\delta_{p_{j^{u-1}_1}p_{j^u_1}}\left(\bar{\Omega}_{u-1}\right)\:\delta_{p_{j^u_1}p_{j^{u+1}_1}}\left(\bar{\Omega}_u\right)\:\delta_{p_{j^{u+1}_1}p_{j^{u+2}_1}}\left(\bar{\Omega}_{u+1}\right) \cdots
\delta_{p_{j^{R}_1}p_1}\left(\bar{\Omega}_{R}\right) = \nonumber \\
(\cdots)\: \delta_{p_1p_{j^u_1}}\left(\bar{\Omega}_1\cap\bar{\Omega}_2\cdots\cap\bar{\Omega}_{u-1}\right)\:  
\delta_{p_{j^u_1}p_{j^{u+1}_1}}\left(\bar{\Omega}_u\right)\:
\delta_{p_{j^{u+1}_1}p_1} \left(\bar{\Omega}_{u+1}\cap\bar{\Omega}_{u+2}\cdots\cap\bar{\Omega}_R\right) = \nonumber\\
(\cdots)\: \delta_{p_{j^u_1}p_{j^{u+1}_1}}\left(\bar{\Omega}_u\cup\left(\bar{\Omega}_1\cdots\cap\bar{\Omega}_{u-1}
 \cap\bar{\Omega}_{u+1}\cdots\cap\bar{\Omega}_{R}\right)\right). \label{eq_joint} 
\end{eqnarray}
Since s$_u$ is unique, that is, $\Omega_u \cap  \Omega_{1,\dots,u-1,u+1,\dots,R} = \emptyset$, the set  $\bar{\Omega}_u\cup\left(\bar{\Omega}_1\cdots\cap\bar{\Omega}_{u-1}
 \cap\bar{\Omega}_{u+1}\cdots\cap\bar{\Omega}_{R}\right) = {\cal M}$. 
In other words, the states $|p_{j^{u+1}_1}\rangle$ and $|p_{j^{u}_1}\rangle$ must coincide everywhere; therefore they are identical: $|p_{j^{u+1}_1}\rangle = |p_{j^{u}_1}\rangle$. This means that 
{\em a joint correlator (\ref{eq_joint}) cannot be irreducible.} 

Nevertheless, for a disjoint correlator this does not hold (Fig.\ref{Fig2-strings}f): $$\bar{\Omega}_u\cup\left(\bar{\Omega}_1\cdots\cap\bar{\Omega}_{R}\right) = \bar{\Omega}_u \equiv {\cal M}\backslash \Omega_u \subset {\cal M},$$ and we cannot conclude that $|p_{j^{u+1}_1}\rangle = |p_{j^{u}_1}\rangle$. The only exception is the case when $s_1$ and $s_R$, the substrings at the beginning and the end of $S$, are mutual, and don't have {\em other} mutuals. Then, from the cyclic invariance of the correlator, these substrings concatenate into a unique substring.

So far we established that an irreducible correlator of the $N$th order does not exist, unless (a) the corresponding  maximal cluster $\Omega$ contains at least $N_{\Omega} = \log_2 N$ qubits, or (b) it is disjoint (its string of indices does not contain a unique substring).
This falls short of our goal to find a signature of equilibrium $N$-qubit  entanglement. Nevertheless the situation changes when we turn to the global response of the system.

\section{Global response of the system in the thermodynamic limit}

The global response of the system is determined by the reaction of its {\em total} magnetization to the external perturbation. It is defined through the 
 {\em average} operators, i.e. operators  of the type 
\begin{equation}S = M^{-1}\sum_{j=1}^M S_j, \end{equation}
and contains in particular the irreducible correlators of  average  operators: 
\begin{equation}
{\cal C}^{\rm irr}_{N;p_1p_2\dots p_N} = M^{-N}\sum_{k_1,k_2,\dots,k_N=1}^{M}c_N^{\rm irr}(k_1,k_2,\dots,k_N; p_1,p_2,\dots, p_N). \label{eq_global}
\end{equation}
The latter are expressed through the irreducible correlators we discussed in the previous Section. 

We will show that in the thermodynamic limit, when both the number of qubits $M$ and the number of maximal clusters $Q$ $\to \infty$, while the order of entanglement $N$ stays finite, the irreducible global correlator ${\cal C}^{\rm irr}_{N}$ (\ref{eq_global}) is zero, unless a finite ratio of maximal clusters in the system contain $\geq N$ qubits each. This makes ${\cal C}^{\rm irr}_{N}$ a signature of  $N$-qubit  equilibrium entanglement. 

In the previous section we saw that there are two types of irreducible correlators, which do not disappear, if the system does not contain $N$-qubit maximal clusters. We will deal with them separately. Without loss of generality, assume that   $M > N$ and $Q > N$.

\subsection{Contributions to  from correlators with repeated indices from the same maximal cluster}

The number of all possible combinations of $N$ indices, each running from 1 to $M$, is $M^N$. The number of combinations of {\em different} indices is $M(M-1)...(M-N+1)$, and the number of combinations with repeating indices is $M^N-M(M-1)...(M-N+1) = O(M^{N-1})$. Therefore in the expression (\ref{eq_global})  the contribution from the correlators with {\em repeating} indices scales as $O(1/M)$.

\subsection{Disjoint correlators}

There are $\left(\begin{array}{c} N-1 \\ R-1 \end{array} \right)$ ways to split a string of $N$ indices in $R$ nonzero substrings. For each substring there are $Q$ maximal clusters we can choose from, and there are on average $(M/Q)^{N/R}$ possible combinations of indices within a cluster. Therefore the total number of different combinations of indices, counted this way, is 
$(M/Q)^N \sum_{R=1}^N \left(\begin{array}{c} N-1 \\ R-1 \end{array} \right) Q^R$.
On the other hand, the number of combinations corresponding to {\em joint} correlators is no less than $(M/Q)^N \sum_{R=1}^N \left(\begin{array}{c} N-1 \\ R-1 \end{array} \right) Q^{R-1}(Q-(R-1))$. Therefore the number of disjoint combinations does not exceed
\begin{eqnarray}
\left(\frac{M}{Q}\right)^N \sum_{R=1}^N \left(\begin{array}{c} N-1 \\ R-1 \end{array} \right) \left(Q^R - Q^{R-1}(Q-(R-1))\right) = 
\left(\frac{M}{Q}\right)^N Q \frac{d}{dQ} \sum_{R=1}^N \left(\begin{array}{c} N-1 \nonumber\\ R-1 \end{array} \right) Q^{R-1} =  \nonumber\\
\left(\frac{M}{Q}\right)^N Q \frac{d}{dQ}\left[(Q+1)^{N-1}\right] = M^N Q^{-1}(N-1)(1+Q^{-1})^{N-2},
\end{eqnarray}
 and the disjoint contribution to the average irreducible correlator (\ref{eq_global}) asymptotically disappears at least as fast as $O(1/Q)$.


\section{Nonlinear susceptibility as a signature of entanglement}

The correlators like (\ref{eq_global})  enter the expressions for (nonlinear) susceptibilities. Therefore  the latter could be used as {\em directly observable} entanglement signatures. We introduce the equilibrium $N$-th order signature,
\begin{eqnarray}
\bar{\cal Z}_N = \left\langle{\cal Z}^n_N\right\rangle \equiv \sum_n{\cal Z}^n_Ne^{-E_n/T},  \label{eq_Z}
\end{eqnarray}
which is the thermal average of the contributions, ${\cal Z}^n_N$, from the sysem's eigenstates $\nr$ with energies $E_n$:
\begin{eqnarray}
 {\cal Z}^n_N = \sum_{p_1}^{'}\sum_{p_2\neq p_1}^{'}\cdots\sum_{p_{N-1}\neq p_{N-2},p_{N-1},\dots p_1}^{'} \frac{M^{-N}{\tt Re}\;\left({\mu^z}_{np_1}{\mu^z}_{p_1p_2}\cdots{\mu^z}_{p_{N-1}n}\right)}{(E_{p_1}-E_n)(E_{p_2}-E_n)\cdots(E_{p_{N-1}}-E_n)}.  \label{eq_Z2}\\ \nonumber
\end{eqnarray}
Here $\mu^z_{pq} = \Pl\mu^z\qr$ is the matrix element of the $z$-component of the system's total magnetic moment. 
The functions $\bar{\cal Z}_N,\:\:{\cal Z}^n_N$ are related to the irreducible part of static nonlinear susceptibility of the system. In the Appendix, this is explicitly shown for the quadratic susceptibility. 

The largest solid state structure where entangled states were demonstrated so far contains four superconducting qubits\cite{Grajcar2005}. This is far from the region of applicability of the thermodynamic limit results of the previous section. Still, a comparison with the experimental data can indicate whether our approach is usable.

We will therefore compare the irreducible part of static ($\omega\to 0$) {\em cubic} susceptiblity, $\chi^{(3)}(\omega) \propto \bar{\cal Z}_4$, to the measure of global entanglement introduced in Ref.\onlinecite{Love} and calculated for the system investigated in Ref.\onlinecite{Grajcar2005}.

The global entanglement measure for an arbitrary pure state $\psi$ of an $n$-qubit system is given by\cite{Love} 
\begin{equation} 
{\cal R}(\psi) = \left(\prod_{1\leq|S|\leq\bar{S}|}' \eta_S(\psi)\right)^{1/(2^{n-1}-1)}. \label{eq_R}
\end{equation}
Here the product is taken over all possible bipartitions of the system, ($S$, $\bar{S}$); $|S|$ is the number of qubits in the set $S$, and the prime denotes that if $|S|=n/2$, we still include every variant only once. The quantity $\eta_S(\psi)$, $0\leq\eta_S\leq 1$ characterizes the entanglement of the set $S$ with the rest of the system, $\bar{S}$  (see Eq.(3) of Ref.\onlinecite{Love}):
\begin{equation}
\eta_S = \frac{2^{|S|}}{2^{|S|}-1} \left\{1 - {\rm tr}\left[\left({\rm tr}_{\bar{S}}\hat{\rho}\right)^2\right]\right\},
\end{equation}
where $\hat{\rho}$ is the density matrix. Obviously $\eta_S = 1$ if and only if  $S$ and $\bar{S}$ are separable. The known Meyer-Wallach\cite{MW} and Scott\cite{Scott} measures of entanglement are expressed through {\em arithmetic} averages of $\eta_S$'s. Unlike them, the measure\cite{Love} in Eq.(\ref{eq_R}) is a {\em geometric} average and therefore is zero if there is even one qubit separable from the rest. 

The expression (\ref{eq_R}) can also be generalized to the case of a mixed state\cite{Love}, but we are here content to use the definition (\ref{eq_R}) and compare the results with the ground-state value of the $4$-th order signature for four qubits,
\begin{equation}
{\cal Z}^0_4 = \sum_{p_1}^{'}\sum_{p_2\neq p_1}^{'}\sum_{p_{3}\neq p_2,p_1}^{'} \frac{4^{-4}{\tt Re}\;\left({\mu^z}_{0p_1}{\mu^z}_{p_1p_2}{\mu^z}_{p_2p_3}{\mu^z}_{p_30}\right)}{(E_{p_1}-E_0)(E_{p_2}-E_0)(E_{p_3}-E_0)}, \label{eq_Z4}
\end{equation}
which is proportional to the irreducible part of the static cubic susceptibility of the system at zero temperature.
Since, unlike ${\cal R}(\psi)$, ${\cal Z}^0_4$ can take negative values, we will take for comparison its absolute value.

The system in question is described by the pseudospin Hamiltonian
\begin{equation}
H = -\sum_{i=1}^4	[\epsilon_i \sigma_z^{(i)} + \Delta_i \sigma_x^{(i)}] +
\sum_{1\leq i<j\leq 4} J_{ij}\sigma_z^{(i)}\sigma_z^{(j)}, \label{eq_Hspin}
\end{equation}
where $\Delta_1 = 147$, $\Delta_2 = 12$, $\Delta_3 = 163$, $\Delta_4 = 165$,
and $J_{12} = J_{34} = 163$, $J_{14} = J_{23} = 155$, $J_{13} = J_{24} = -62$ (all in mK) were determined from the experiment\cite{Grajcar2005}. The operator $\mu_z \propto \sum_{i=1}^{4}\sigma_z^{(i)},$ and all the terms in Eq.(\ref{eq_Z4}) can be calculated explicitly. 

\begin{figure}[t]
\includegraphics[width=8cm]{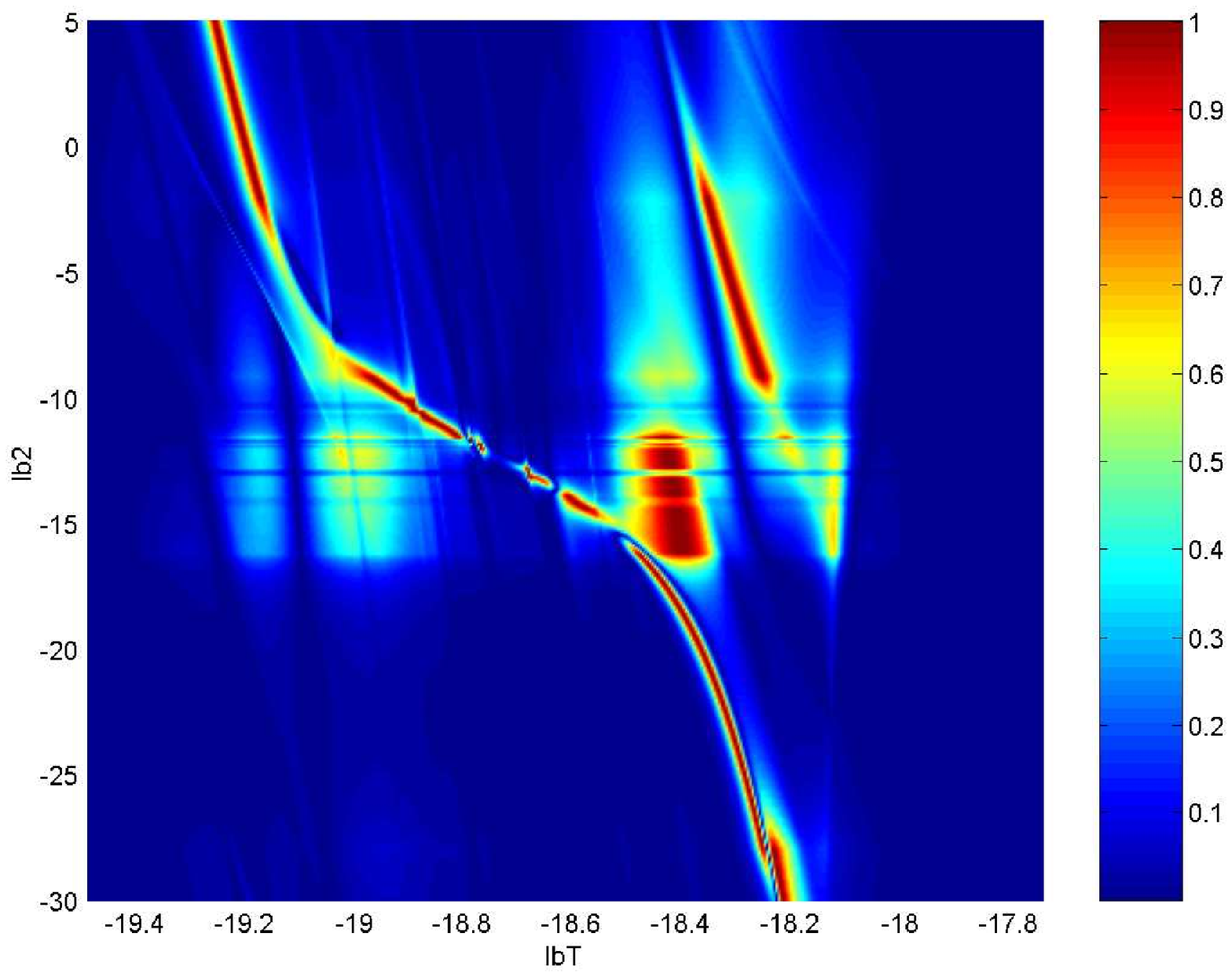} 
\includegraphics[width=8cm]{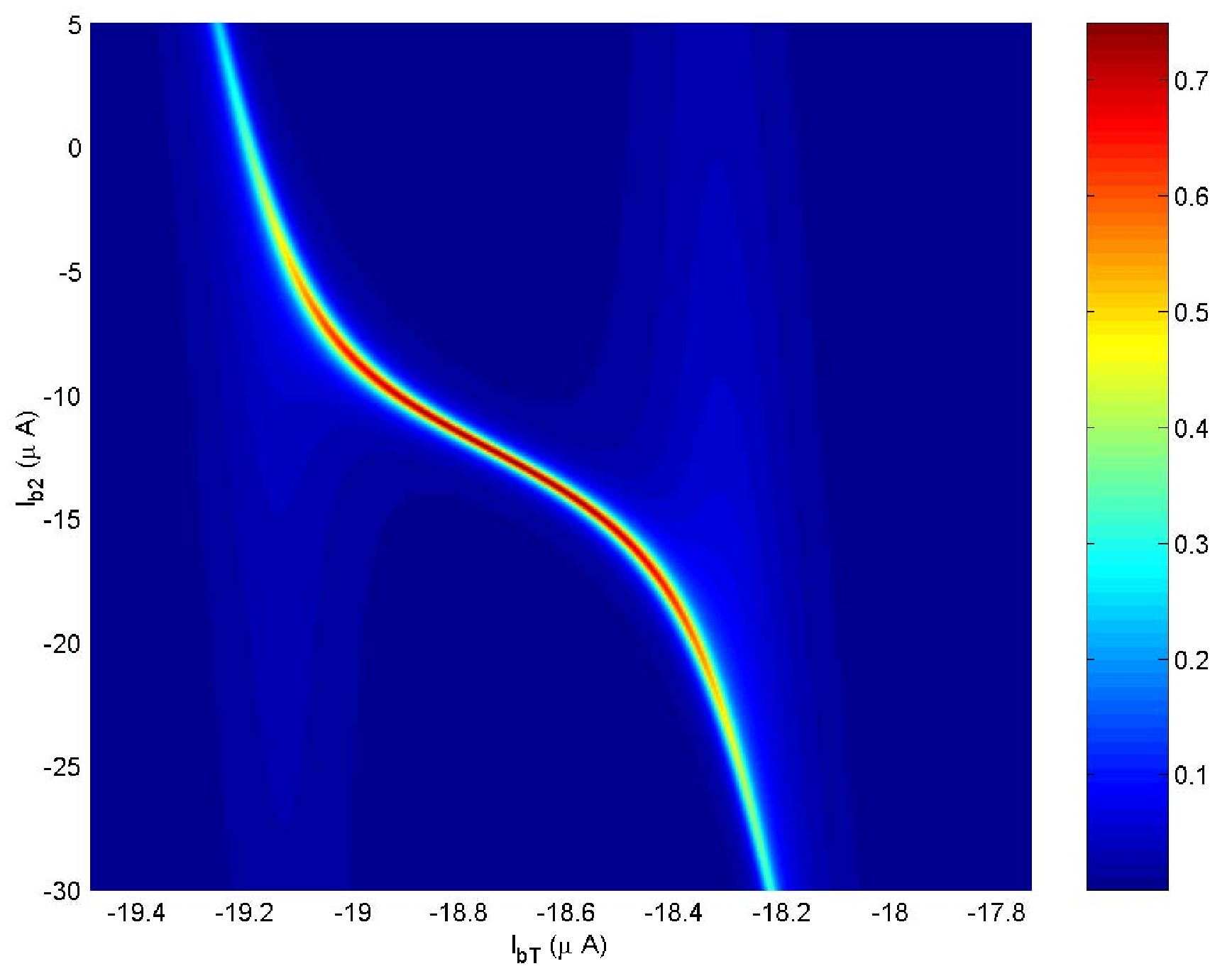}\\
 \caption{(Left) Normalized 4-qubit ground-state entanglement signature  $|{\cal Z}^0_4(I_{b2},I_{bT})|/\max_{I_{bT}}|{\cal Z}^0_4(I_{b2},I_{bT})|$ (Eq.(\ref{eq_Z4})),
  and (right) global 4-qubit entanglement measure  ${\cal R}(\psi_0)$ (Eq.(\ref{eq_R})), as a function of bias currents in the 4-flux-qubit system of Ref.\onlinecite{Grajcar2005}.   }\label{Fig-top view}
\end{figure}

The dependence of $|{\cal Z}^0_4|$ and ${\cal R}(\psi_0)$ (ground state global entanglement) are shown in Fig.\ref{Fig-top view} as a function of variables $I_{bT}$ and $I_{b2}$. These variables correspond to the bias currents through different tuning coils in the actual experiment and can be explicitly expressed in terms of the biases $\epsilon_i$ in Eq.(\ref{eq_Hspin}) (see Ref.\onlinecite{Grajcar2005}).  Fig.\ref{Fig-top view} represents one   possible section  of the parameter space $\{\epsilon_i\}_{i=1}^{4}$. The centre of this picture is the co-degeneracy point, i.e. here all qubits in (\ref{eq_Hspin}) are biased to their degeneracy. The maximum value of $\cal R$ for four qubits is approximately 0.95 (Ref.\onlinecite{Love}). To make scales comparable, we normalized $|{\cal Z}^0_4|$ to its maximum value at a given $I_{b2}$.

\begin{figure}[hb]
\includegraphics[width=8cm]{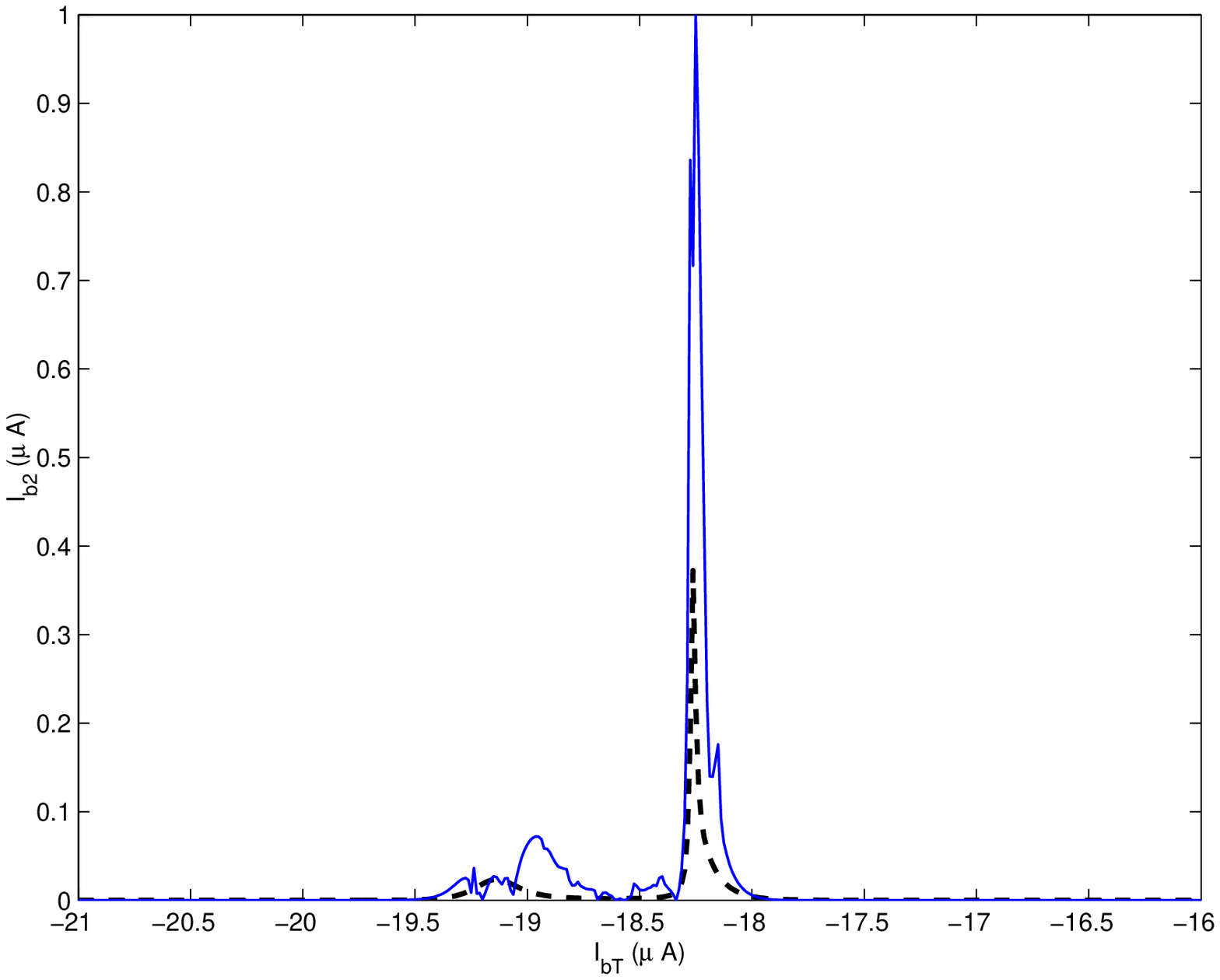} 
\includegraphics[width=8cm]{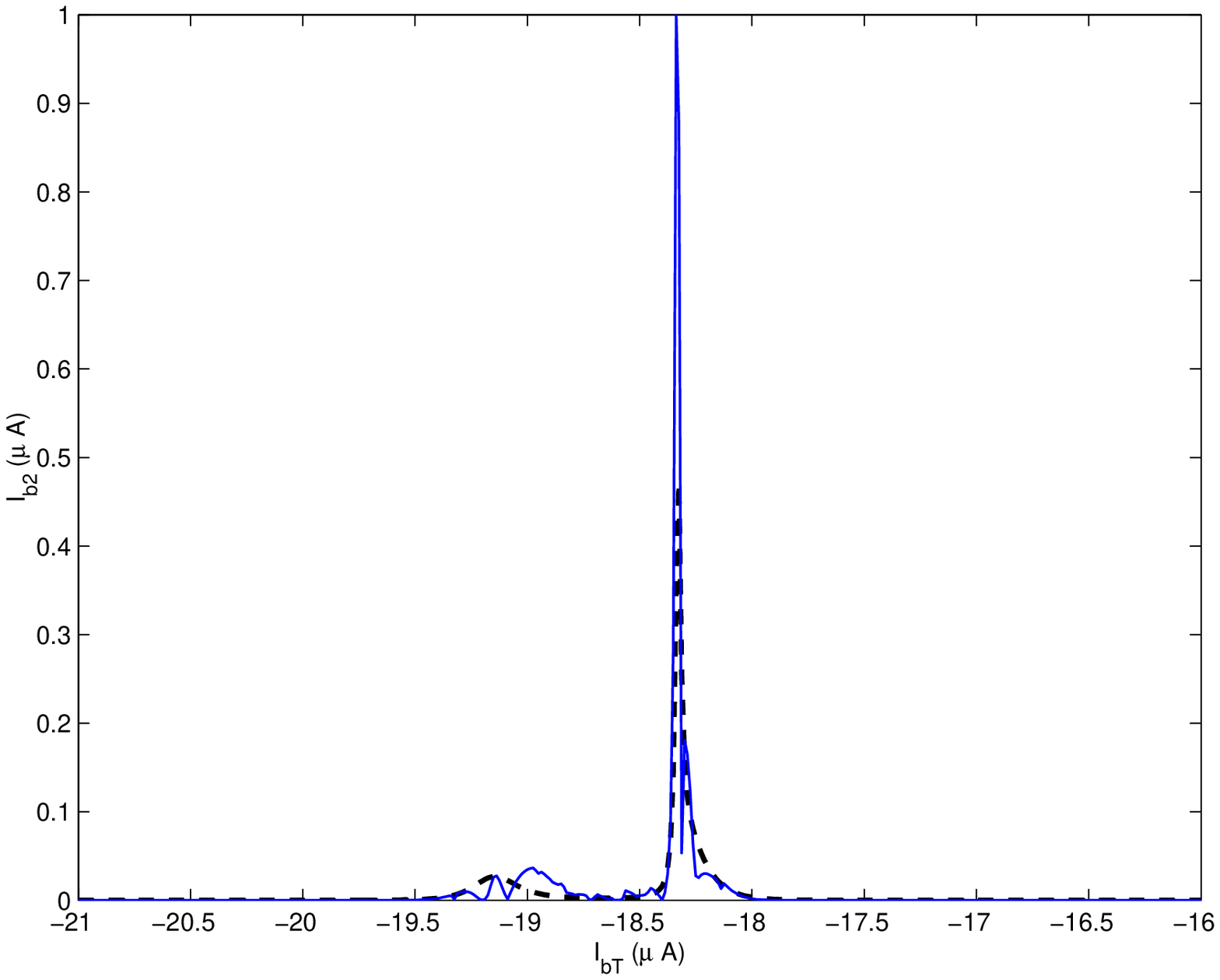}\\
\includegraphics[width=8cm]{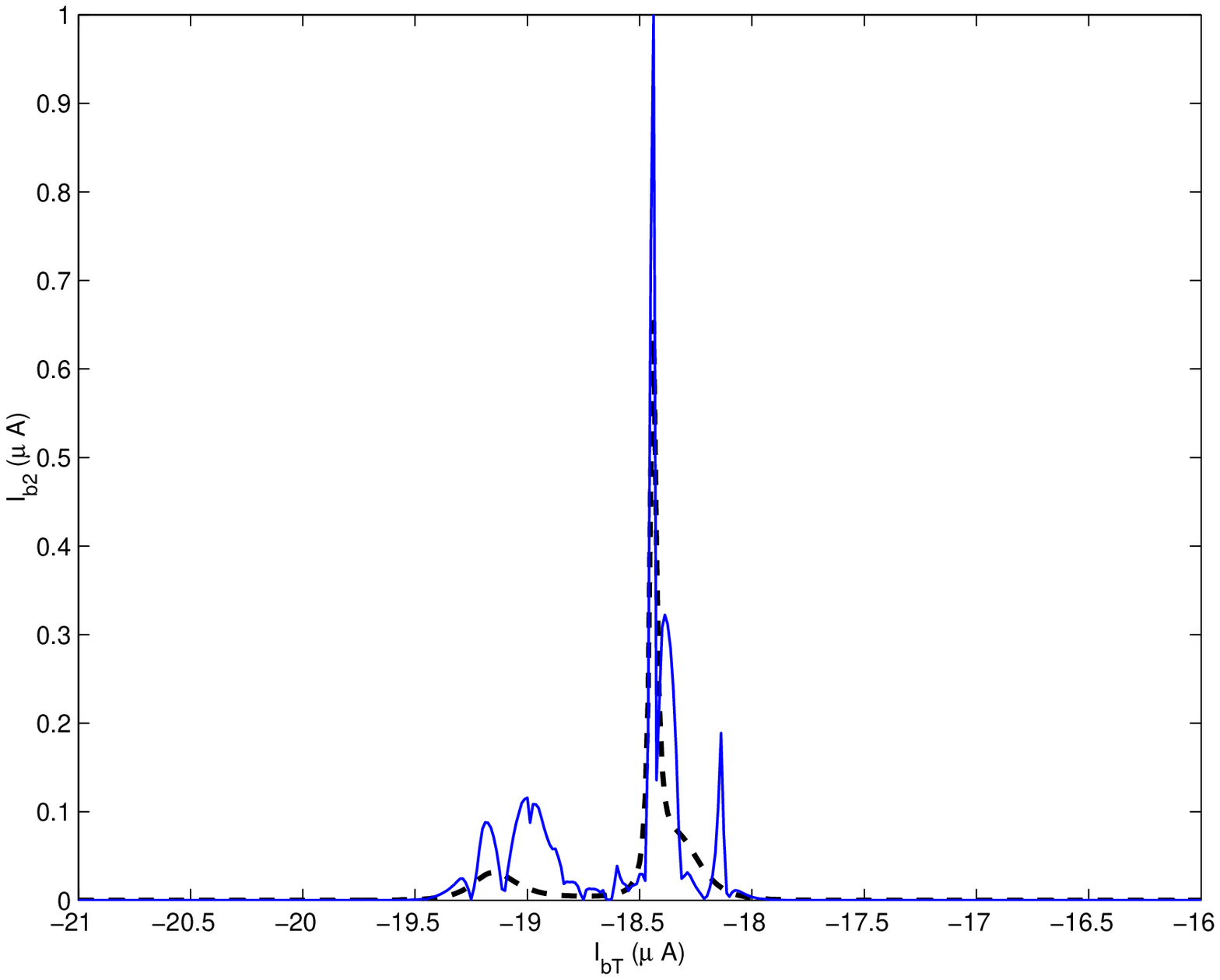} 
\includegraphics[width=8cm]{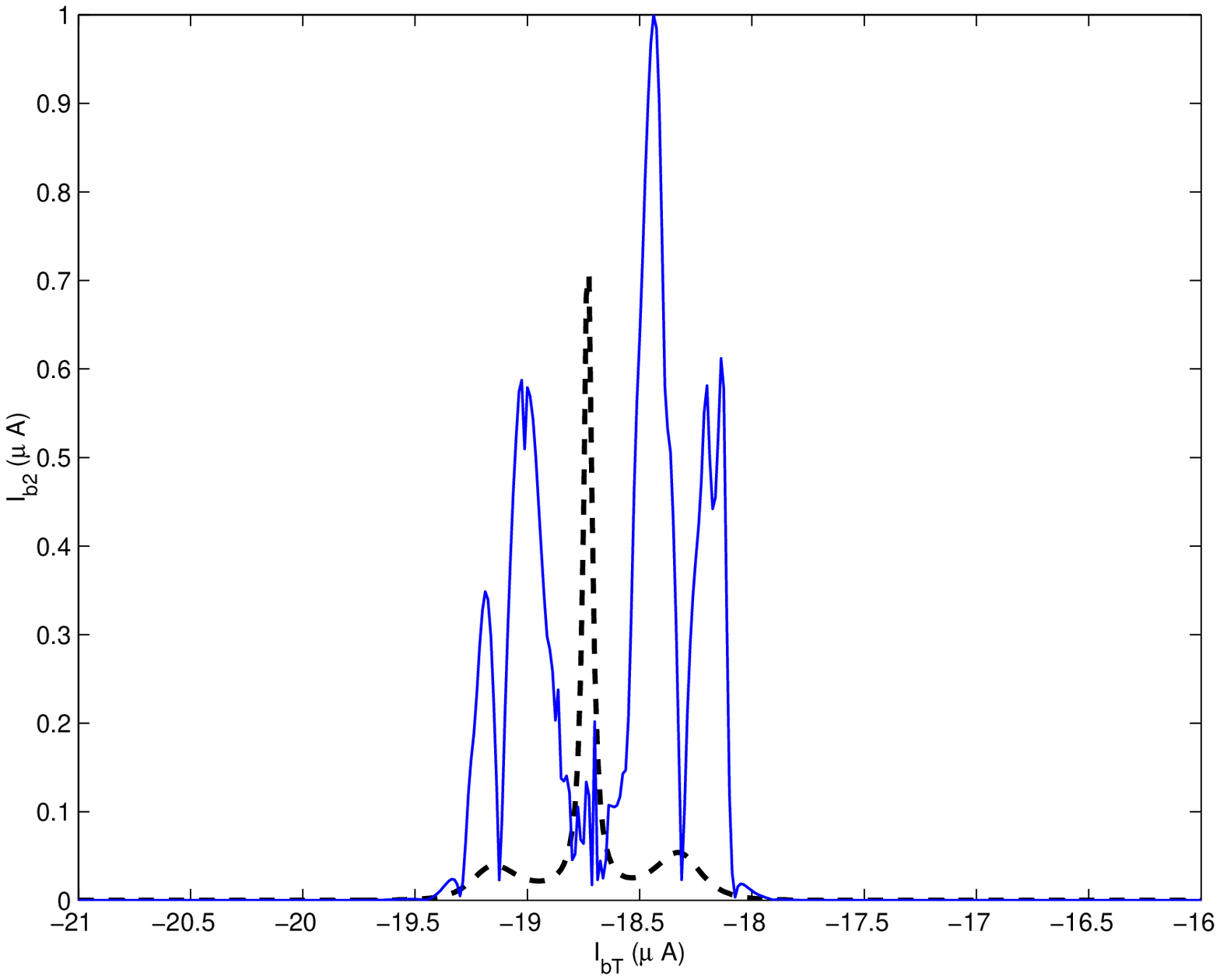}\\ 
\includegraphics[width=8cm]{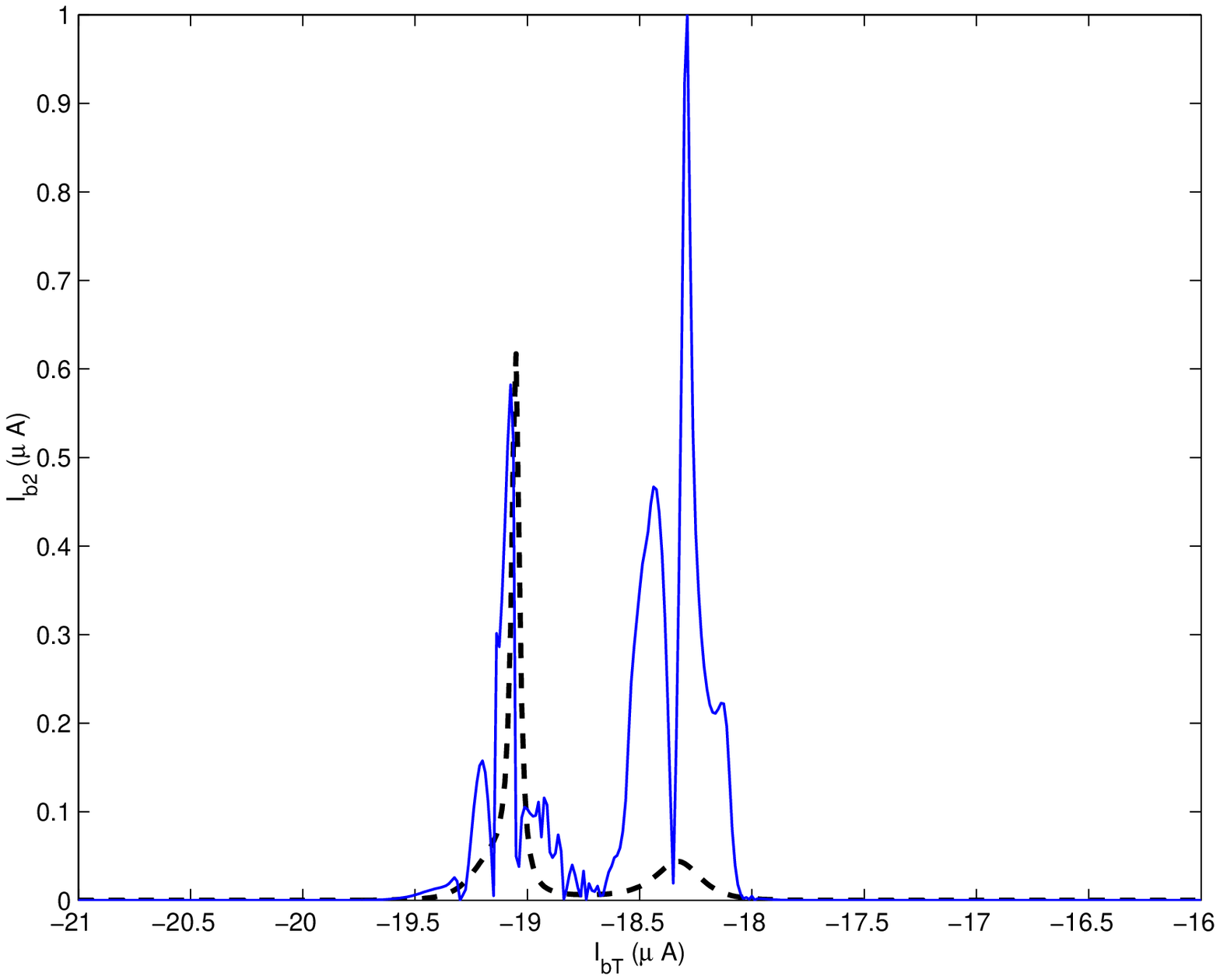}
\includegraphics[width=8cm]{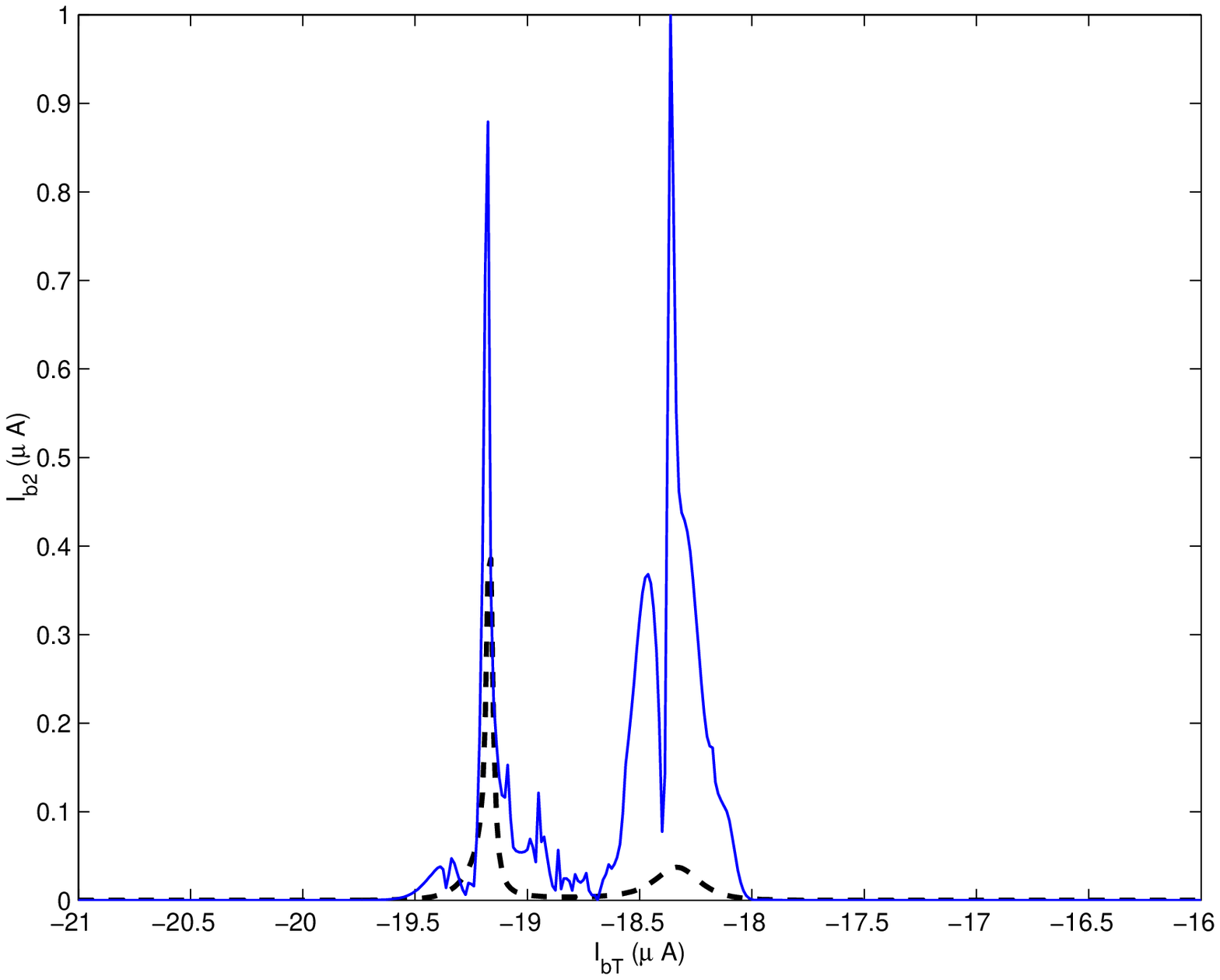}\\ 
\includegraphics[width=8cm]{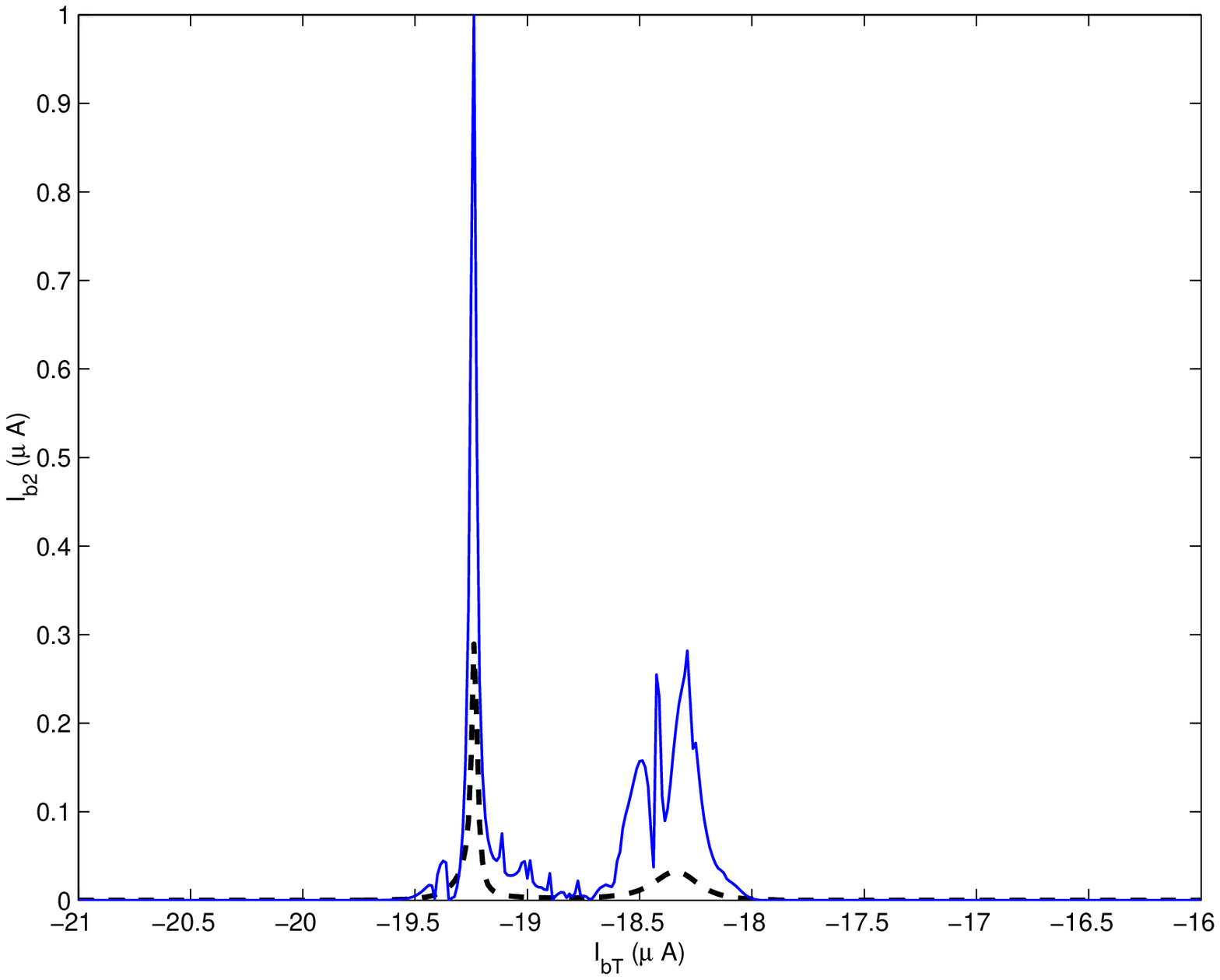}
\includegraphics[width=8cm]{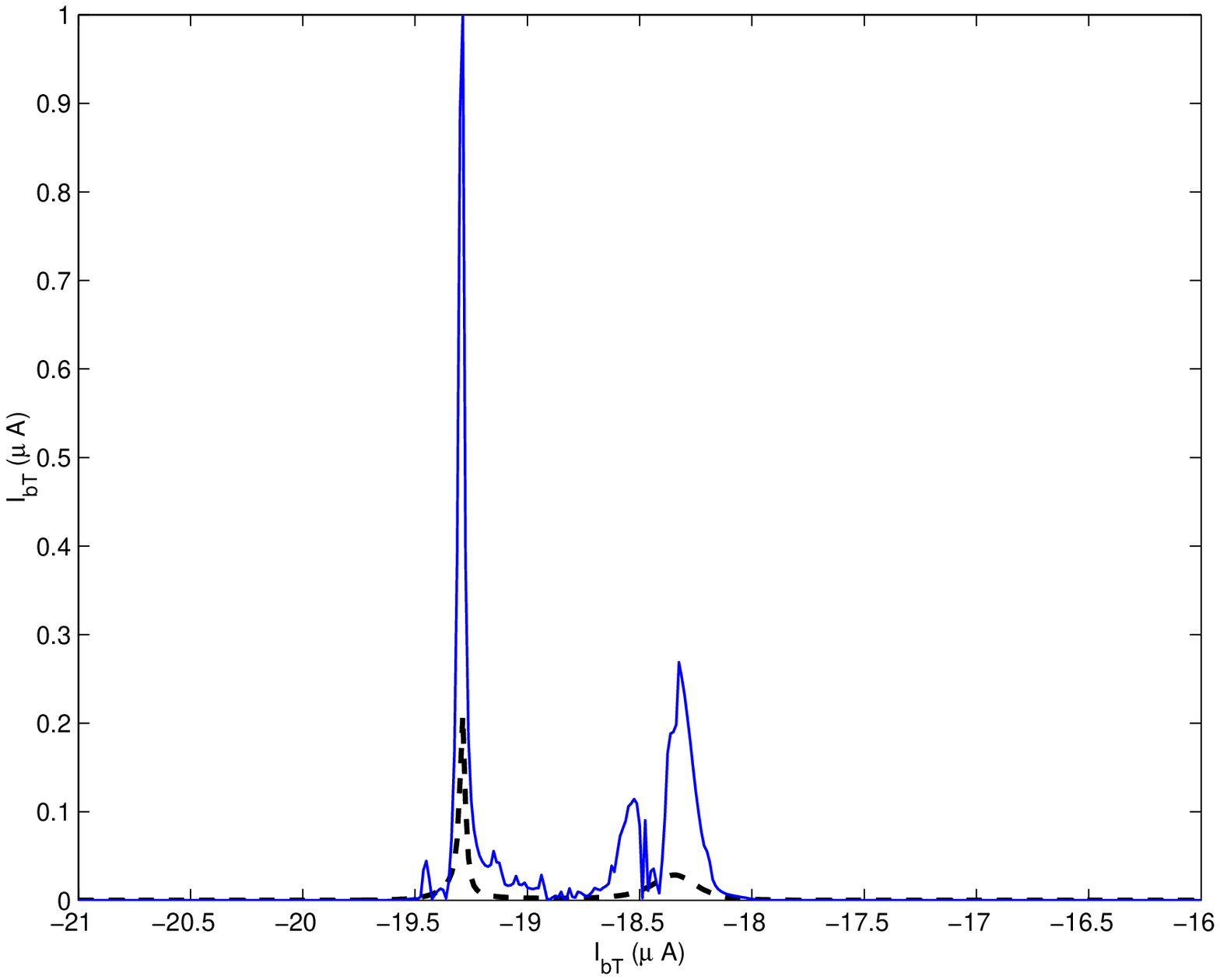}\\
 \caption{(Colour online.) Normalized 4-qubit entanglement signature  
 $|{\cal Z}^0_4(I_{b2},I_{bT})|/\max_{I_{bT}}|{\cal Z}^0_4(I_{b2},I_{bT})|$ 
 (solid line) and global 4-qubit entanglement measure  ${\cal R}(\psi_0)$ (dotted line). The values of the bias current $I_{b2}$ (all in $\mu$A) are 
 (left to right, top to bottom) -27.5; -22.5; -17.5; -12.5;  -7.5; -2.5; 2.5; 7.5.}
 \label{Fig-section view}
\end{figure}

One can immediately see that the structure and positions of the maxima of both functions practically coincide, though the relative heights of the peaks not necessarily so.   This is further illustrated in Fig.\ref{Fig-section view}, where the dependences on $I_{bT}$ for a given value of $I_{b2}$ are presented.   Given that the formulae like (\ref{eq_Z4}) are only supposed to hold in the thermodynamic limit, such a similarity between the behaviour of $\cal R$ and ${\cal Z}^0_4$ is  remarkable and indicates that the irreducible part of nonlinear susceptibility can provide information about multiqubit  entanglement even in modest-sized structures available now.

\section{Conclusions}

In this paper we concentrated on equilibrium entanglement, that is, on the presence of entangled $N$-qubit clusters in the eigenstates of the Hamiltonian in a multiqubit system, and its directly observable manifestations.
We demonstrated that under plausible enough assumptions (Russian-doll condition), and in the thermodynamic limit, the irreducible part of static nonlinear susceptibility provides a signature of the existence of multiqubit  entanglement in the system. We introduced a series of such entanglement signatures. We also considered a four-qubit system, with the choice of parameters corresponding to the experimentally realized situation. We saw that 
the structures of maxima of the global four-qubit entanglement in the ground state, $\cal R$, and of the fourth-order entanglement signature, ${\cal Z}^0_4$, as a function of control parameters, are almost identical. This raises an expectation that 
the method we introduced can provide a directly observable evidence of multiqubit  entanglement even in systems of moderate size.

There  remain open questions. First, it is necessary to investigate the requirements to the Hamiltonians which satisfy the Russian-doll condition.  Second, one would like to see explicitly the origin of similarity between $\cal R$ and $\cal Z$, and investigate whether one can be expressed through the other in some nontrivial limit. Third, it is desirable to develop an experimental protocol, which would allow to extract $\cal Z$ of the third or fourth order from the measurements and thereby provide a direct evidence of equlibrium multiqubit  entanglement.

\section{Acknowledgments}

We are grateful to I. Affleck, M. Grajcar, Y. Harada, E. Il'ichev, D. Lidar, A. Maassen van den Brink, M.A. Martin-Delgado, K. Maruyama, F. Nori, V. Petrashov, S. Rashkeev, T. Roscilde, A. Tzalenchouk, and J. Young for helpful discussions at different stages of this work. AZ was partially supported by NSERC Discovery Grant (Canada).

\appendix

\section{Irreducible part of static quadratic susceptibility}

We start with the Liouville equation for the density matrix,
\begin{equation}
i\partial_t \rho(t) = [H_0 + H_1(t), \rho(t)], \label{eq1}
\end{equation}
where the unperturbed Hamiltonian of $M$ flux qubits and the perturbation are
\begin{equation}
H_0 = -\frac{1}{2}\sum_{j=1}^M \left(\D_j \sx_j + \e_j \sz_j\right)
-\sum_{i<j} J_{ij} \sz_i \sz_j; \:\:\:\: H_1(t) = -h(t) \sum_{j=1}^M \left(\l^z_j \sz_j + \l^x_j \sx_j\right). \label{eq_pseudospin}
\end{equation}
 The factors $\l_j$ answer e.g. for the inhomogeneous external magnetic field distribution in the system.  The equilibrium entanglement we are going to probe is created by the unperturbed Hamiltonian, $H_0$. (Of course, the following discussion is applicable to any system described by the pseudospin Hamiltonian (\ref{eq_pseudospin}).)

Iterating Eq.(\ref{eq1}), one obtains the standard expansion of the density matrix over the commutators of $H_1$ and $\rho_0$ (unperturbed density matrix $\rho_0$ is time-independent and commutes with $H_0$, but it does not have to be an equilibrium density matrix for $H_0$). The expansion $\rho(t) = \rho^{(0)} + \rho^{(1)}(t) + \rho^{(2)}(t) +\dots$ leads to the corresponding expansion for the magnetic moment of the system, ${\vec \mu}(t) = \tr \left\{\rho(t)\sum_j {\vec \sigma}_j\right\}.$

One can measure different components of the magnetization and apply several components of the perturbation. 
Here we consider the situation closest to the 2- and 4-qubit setup of \cite{Izmalkov2004,Grajcar2005,Love}, where both the external field, and the measured magnetic moment are along $z$-axis. For simplicity we also assume that all $\l$'s are the same.

The linear term in the expansion of $\mz(t)$ is related to 2-tanglement and was treated in \cite{Izmalkov2004,Smirnov03}. The first new term is $\mz^{(2)}(t)$,
\begin{eqnarray}
\mz^{(2)}(t) =\!\!\!\! \sum_{i,j,k=1}^N\frac{1}{i^2} \int_{-\infty}^{t} \!\!\!\!\!\!dt'\int_{-\infty}^{t'}\!\!\!\!\!\!dt'' h(t')h(t'') \tr\left\{
\rho^{(0)}\left[\left[\sz_i(t),\sz_j(t')\right],\sz_k(t'')\right]\right\} =   \int_{-\infty}^{\infty} \!\!\!\!\!\!dt'\int_{-\infty}^{\infty}\!\!\!\!\!\!dt'' \x2(t-t',t-t'') h(t')h(t''),  
\end{eqnarray}
where the quadratic susceptibility $\x2$ is
\begin{eqnarray}
\x2(t-t',t-t'') = \sum_{i,j,k=1}^N\frac{1}{i^2} \Q(t-t')\Q(t'-t'')
\tr\left\{
\rho^{(0)}\left[\left[\sz_i(t),\sz_j(t')\right],\sz_k(t'')\right]\right\}.
\label{eq2}
\end{eqnarray}

In Fourier components 
we get
\begin{equation}
\mz^{(2)}(t) = \int_{-\infty}^{\infty} \!\!\frac{d\omega}{2\pi}\int_{-\infty}^{\infty}\!\!\frac{d\omega'}{2\pi} e^{-i(\omega+\omega')t}\x2(\omega,\omega') h(\omega) h(\omega').
\end{equation}
For a monochromatic perturbation, $h(\omega)=2\pi {\cal H} \delta(\omega-\omega_0),$ the response is $
\mz^{(2)}(t) = e^{-2i\omega_0t} {\cal H}^2 \x2(\omega_0,\omega_0).
$

We can rewrite (\ref{eq2}) using the eigenstates of $H_0$, $H_0\nr = E_n\nr$,  $n=1,2\dots 2^M$:
\begin{equation}
\x2(t-t',t-t'') = \frac{1}{i^2}\sum_{ijk}\sum_{n=1}^{2^M} \rho^{(0)}_{n} \Q(t-t')\Q(t'-t'')\nl \left[\left[\sz_i(t),\sz_j(t')\right],\sz_k(t'')\right]\nr, 
\end{equation}
 where all operators $\sz(t)$ are in the interaction representation. Opening the commutators, using the closure relations, taking the Fourier transform and doing the usual algebra, we finally obtain:
\begin{equation}
\x2(\omega,\omega') = \sum_{n=1}^{2^M}\rho^{(0)}_n \x2^n(\omega,\omega') 
\equiv \sum_{n=1}^{2^M}\rho^{(0)}_n \left[\sum_{p,q=1}^{2^M} {\cal C}_{n;pq} f_{n;pq}(\omega,\omega')\right].
\end{equation}
We introduced the weight factors 
\begin{equation} 
{\cal C}_{n;pq} = \sum_{ijk} \nl\sz_i\pr\Pl\sz_j\qr\ql\sz_k\nr \equiv 
\nl{\mu^z}\pr\Pl{\mu^z}\qr\ql{\mu^z}\nr, \label{eq_3tan}
\end{equation}
and the formfactors
\begin{eqnarray}
f_{n;pq}(\omega,\omega') = \frac{1}{\omega+\omega'-(E_q-E_p)+i0} \times \label{eq_f} \\
\left\{ \frac{E_n+E_q-2E_p}{(\omega'-(E_q-E_n)+i0)(\omega+\omega'-(E_p-E_n)+i0)} - 
\frac{E_n+E_p-2E_q}{(\omega'-(E_n-E_p)+i0)(\omega+\omega'-(E_n-E_q)+i0)}\right\}.\nonumber
\end{eqnarray} Obviously, ${\cal C}_{n;pq} =  {\mu^z}_{np} {\mu^z}_{pq} {\mu^z}_{qn}  = {\cal C}_{npq} = {\cal C}_{pqn}= {\cal C}_{qnp}$, and $f_{n;nn}(\omega,\omega')=0$.

In the  setup of \cite{Izmalkov2004,Grajcar2005,Love} the frequency of the $LC$-tank, at which the susceptibility is measured, is orders of magnitude less than the tunneling and coupling frequencies ($\sim 20$ MHz vs. $\sim 1$ GHz).  The interlevel spacings are at least $\sim \Delta/M$. Therefore if $\omega \ll \Delta$ we can use the static limit:
$$
\chi_0^n = \lim_{\omega\to 0} \sum_{pq}'' {\cal C}_{npq} f_{n;pq}(\omega,\omega),
$$
where $\sum''$ means that terms with $E_p=E_q=E_n$ are excluded. Separating the contributions with both $E_p \neq E_n, E_q \neq E_n$ yields 
\begin{eqnarray*}
\sum_p'\sum_q' \frac{-3{\cal C}_{npq}}{(E_p-E_n)(E_q-E_n)} =
\sum_p'\frac{-3{\cal C}_{npp}}{(E_p-E_n)^2} - 6\sum_p' \sum_{q<p}'\frac{{\tt Re}\;({\cal C}_{npq})}{(E_p-E_n)(E_q-E_n)}.
\end{eqnarray*}
Here $\sum'$ means that the terms with $E_p=E_n$  are excluded. 
The rest are the terms where either $p=n$ or $q=n$, and since 
$
{\cal C}_{nnq} = {\cal C}_{nqn} = |{\mu^z}_{nq}|^2 {\mu^z}_{nn},
$
they contribute
\begin{eqnarray*}
\sum_{q}' {\cal C}_{nnq} \lim_{\omega\to 0}(f_{n;nq}(\omega,\omega) + f_{n;qn}(\omega,\omega)) = 
\sum_q' \frac{+3 {\cal C}_{nnq}}{(E_n-E_q)^2}.
 \end{eqnarray*}
Bringing together all the terms, we finally obtain the low-frequency limit for the quadratic susceptibility:
\begin{eqnarray}
\chi_0^n =    
3 \sum_q' \frac{ |{\mu^z}_{nq}|^2 ({\mu^z}_{nn}-{\mu^z}_{qq})}{(E_q-E_n)^2} - 6\sum_p' \sum_{q<p}'\frac{{\tt Re}\;({\cal C}_{npq})}{(E_p-E_n)(E_q-E_n)} \equiv \chi_{0A}^n + \chi_{0B}^n . \label{eq_LF_chi}
 \end{eqnarray}

The meaning of the two terms in Eq.(\ref{eq_LF_chi}) is different. Indeed,  consider a system of noninteracting qubits. Then the eigenstates are completely factorized, 
$\pr = \prod_{j=1}^M |p_j\rangle,$ and the state of the $j$th qubit is either $|g_j\rangle$ (ground) or $|e_j\rangle$ (excited), with energy $\epsilon_{g(e),j}$. 
 Denoting by $\bar{n}$ the switched state ($\bar{g} = e, \bar{e} = g$), we  obtain for $\chi_{0A}^n$
\begin{eqnarray}
\chi_{0A, {\rm fact}}^n =   \sum_{i=1}^M \frac{3\left|\langle n_i|\sz_i|\bar{n}_i\rangle\right|^2
\left(\langle n_i|\sz_i|n_i\rangle - \langle\bar{n}_i|\sz_i|\bar{n}_i\rangle\right)}{(\epsilon_{\bar{n},i}-\epsilon_{n,i})^2}, \label{eq_chiA_factor}
\end{eqnarray}
that is, the sum of single-qubit susceptibilities.  On the other hand, one can check that 
\begin{equation}
\chi_{0B}^n  
 \equiv - 6\sum_p' \sum_{q<p}'\frac{{\tt Re}\;({\mu^z}_{np} {\mu^z}_{pq} {\mu^z}_{qn} )}{(E_p-E_n)(E_q-E_n)}
\end{equation}
 is exactly zero for the system of independent qubits. Moreover, the summation restrictions in (\ref{eq_LF_chi}) mean that the coefficients ${\cal C}_{npq}$ can be replaced by the irreducible correlators (\ref{eq_global}) of the operators $\sz_i$:
\begin{equation}
{\cal C}_{npq} \to M^3 {\cal C}^{\rm irr}_{3; npq} = \sum_{i,j,k} c_3^{\rm irr}(i,j,k; n,p,q). \label{eq_Cc}
\end{equation}
Therefore the function $\chi_{0B}/(M^3)$ will disappear in the thermodynamic limit in the absence of  entangled 3-qubit clusters. Thus it provides an observable signature of 3-tanglement\cite{chiB}. (Note that for $N=3$ the contribution from the disjoint correlators is exactly zero, due to cyclic invariance.)
This signature can be extracted from the measurements by detecting the  second harmonic generation when one, two etc qubits are near the co-degeneracy point and comparing the outcomes, much like it was done with the IMT deficit of \cite{Izmalkov2004}.

Now we can introduce the $N$-th order signature, Eq.(\ref{eq_Z}).
 The functions $\bar{\cal Z}_N$ are related to directly observable global nonlinear response of the system: the measurable quantity $\langle\chi_{0B}\rangle$\cite{chiB}  is thus proportional to $\bar{\cal Z}_3$ the same way the observed IMT deficit of Ref.\cite{Izmalkov2004} is proportional to $\bar{\cal Z}_2$. We saw that the ${\cal Z}_N$-functions are asymptotically zero in the thermodynamic limit, unless the system contains a  finite proportion of entangled clusters of at least $N$ qubits each; they are also directly related to the irreducible contribution to the corresponding static nonlinear susceptibilities.  Therefore the functions (\ref{eq_Z},\ref{eq_Z2}) can indeed serve as observable entanglement signatures.


\begin{thebibliography}{99}
\bibitem{chiB} Strictly speaking, what can be measured is
$
\langle\chi_{0B}\rangle =  \sum_{n=0}^{2^M-1}\rho_n \chi_{0B}^n,
$
the average over the stationary state of the system, but for low enough effective temperature we'll be close to the ground state value, $\chi_{0B}^0$.
\end{thebibliography}
\end{document}